\documentclass[12pt]{article}
\usepackage{amsmath,amssymb}
\usepackage[hypertex]{hyperref}
\usepackage{graphicx,color}

\def\bb#1{\mathbb{#1}}
\def\pare#1{\left( #1\right)}
\def\bpare#1{\left\{ #1\right\}}
\def\cpare#1{\left[ #1\right]}
\def\defeq{\,\raisebox{0.4pt}{:}\hspace{-1.2mm}=}
\def\ds{\displaystyle}
\def\ssp{\hspace{0.3mm}}
\def\sn{\,\mathrm{sn}}
\def\cn{\,\mathrm{cn}}
\def\dn{\,\mathrm{dn}}
\def\iom{i\hspace{0.2mm}\omega}
\def\iomm#1{i\hspace{0.2mm}\omega_{#1}}
\def\Re{\mathop{\rm Re}\nolimits\,}
\def\Im{\mathop{\rm Im}\nolimits\,}
\def\eK{\mathrm{\bf K}}
\def\eE{\mathrm{\bf E}}
\renewcommand{\eqref}[1]{$\pare{\rm \ref{#1}}$}

\newcommand{\f}[2]{\frac{#1}{#2}}
\newcommand{\ko}[1]{\left( #1 \right)}
\newcommand{\kko}[1]{\left[ #1 \right]}
\newcommand{\kkko}[1]{\left\{ #1 \right\}}
\newcommand{\abs}[1]{\left| #1 \right|}

\newcommand{\komoji}[1]{\mbox{$#1$}}

\def\cN{{\mathcal{N}}}
\def\cL{{\mathcal{L}}}
\def\cT{{\mathcal{T}}}
\def\cJ{{\mathcal{J}}}
\def\cE{{\mathcal{E}}}
\def\pa{\partial}
\def\eq{\equiv}

\def\sig{\sigma}
\def\al{\alpha}

\def\om{\omega}

\def\lam{\lambda}
\def\tlam{\widetilde\lambda}
\def\Th{\Theta}

\def\AdSS{{AdS${}_5\times {}$S${}^5$}}
\def\S{{S${}^5$}}
\def\RS{{$\mathbb{R}_{\rm t} \! \times {\rm S}^3$}}
\newcommand{\SN}[2]{\,\mathrm{sn}\left( #1 , #2 \right)}
\newcommand{\CN}[2]{\,\mathrm{cn}\left( #1 , #2 \right)}
\newcommand{\DN}[2]{\,\mathrm{dn}\left( #1 , #2 \right)}
\newcommand{\mr}[1]{\mathrm{#1}}
\numberwithin{equation}{section}

\begin{document}

\vspace{-1.5cm}
\begin{flushright}
\parbox{3.5cm}
{
{\bf September 2006}\hfill\\
UT-06-18 \hfill\\ 
{\tt hep-th/0609026}\hfill 
}
\end{flushright}

\vspace*{0.5cm}

\begin{center}
\Large\bf
A Perspective on Classical Strings \\[4mm]
from Complex Sine-Gordon Solitons
\end{center}
\vspace*{0.7cm}
\centerline{\large 
Keisuke Okamura$^{a}$ 
~and~
Ryo Suzuki$^{b}$}
\vspace*{0.5cm}
\begin{center}
\emph{Department of Physics, Faculty of Science, University of Tokyo,\\
Bunkyo-ku, Tokyo 113-0033, Japan.} \\
\vspace*{0.5cm}
${}^{a}${\tt okamura@hep-th.phys.s.u-tokyo.ac.jp}\\
${}^{b}${\tt ryo@hep-th.phys.s.u-tokyo.ac.jp}
\end{center}

\vspace*{0.7cm}

\centerline{\bf Abstract} 

\vspace*{0.5cm}

We study a family of classical string solutions with large spins on \RS\ subspace of {\AdSS} background, which are related to Complex sine-Gordon solitons via Pohlmeyer's reduction.
The equations of motion for the classical strings are cast into Lam\'{e} equations and Complex sine-Gordon equations.
We solve them under periodic boundary conditions, and obtain analytic profiles for the closed strings.
They interpolate two kinds of known rigid configurations with two spins\,: 
on one hand, they reduce to folded or circular spinning/rotating strings in the limit where a soliton velocity goes to zero, while on the other hand, the dyonic giant magnons are reproduced in the limit where the period of a kink-array goes to infinity.

\vfill

\thispagestyle{empty}
\setcounter{page}{0}

\newpage

\section{Introduction\label{sec:intro}}

The AdS/CFT correspondence \cite{Maldacena97,GKP98,Witten98} predicts a remarkable matching between string and gauge theories.
The best studied example of it is the one between string theory on {\AdSS} and four-dimensional $\cN=4$ super Yang-Mills (SYM) theory.
In recent years, there have been significant progress not only in BPS but also in non-BPS sectors, and many non-trivial tests has been carried out by virtue of integrabilities of both sides.
Semiclassical spinning/rotating string solutions with energies analytic in the effective coupling $\tlam\eq \lam/J^{2}$, which is the 't Hooft coupling divided by the large-spin squared, were shown to be dual to ``long'' composite operators of the SYM theory side in the limit $\lam\to \infty$ while $\lam/J^{2}\ll 1$ kept fixed.
In the dual gauge theory side, $J$ represents the R-charge of a SYM operator.
This limit is known as Berenstein-Maldacena-Nastase (BMN) limit \cite{BMN02}, and in this limit the worldsheet momentum $p\sim 1/J$ goes to zero.
In both near-BPS (BMN) and far-from-BPS sectors, the string energy $E(J;\lam)$ and the anomalous dimension $\Delta(J;\lam)$ of SYM operators are expanded in powers of the effective coupling as $E=J+c_{1}\tlam+c_{2}\tlam^{2} + \ldots$ and $\Delta=J+a_{1}\tlam+a_{2}\tlam^{2} + \ldots$, which enabled us to test the key proposal of the AdS/CFT quantitatively, that is, to check $a_{k}\stackrel{?}{=}c_{k}$ $(k=1,2,\dots)$.

Various types of string configurations were studied in this context \cite{GKP02, FT03b, FT03c, FT03, Minahan02, AFRT03},\footnote{\,For review articles, see \cite{Tseytlin03,Tseytlin04}.} and were compared to their gauge theory duals.
Among them, the correspondence between $(i)$ so-called (elliptic) folded strings and ``double contour'' configurations of Bethe roots in the gauge theory spin-chain, and $(ii)$ so-called circular string and ``imaginary root'' configurations of Bethe roots, provided nice examples in checking the duality at the level of concrete solutions \cite{BMSZ03,BFST03}.
Perturbative expansion of their energies for both $(i)$ and $(ii)$ cases revealed a remarkable agreement with the SYM counterparts including and up to the two-loop level, {\it i.e.}, $a_{1}=c_{1}$ and $a_{2}=c_{2}$, in quite a non-trivial fashion.
At the three-loop level, however, the coefficients turn out to disagree, $a_{3}\neq c_{3}$, which is known as the ``three-loop discrepancy'' \cite{SS04, CLMSSW03, CMS04, RSS05}. It still remains a challenging problem.

Recently, Hofman and Maldacena (HM) considered another interesting limit \cite{HM06}.
As compared to the BMN limit, this time $J$ is again taken to infinity, whereas $\lam$ can be kept finite and $p$ is also fixed.
In the HM limit, both strings and dual spin-chains become infinitely long, and both sides of the correspondence are characterized by a centrally-extended SUSY algebra $\mr{SU(2|2)}\times \mr{SU(2|2)} \ltimes {\mathbb R}^{2}$.
``Asymptotic" spin-chain states for an elementary magnon case were considered in \cite{Beisert05} to determine the central charge of the $\mr{SU(2|3)}$ dynamic spin-chain state, and it was generalized to magnon bound states case in \cite{Dorey06}.
Let $Q$ be the number of constituent magnons for the bound state in the infinite spin-chain, then the BPS condition of the extended SUSY algebra determines the dispersion relation to be
\begin{equation}
\Delta-J_{1}=\sqrt{Q^{2}+ f(\lambda) \sin^{2}\ko{\f{p}{2}}}\qquad 
(\Delta\,,\,J_{1}\to \infty)\,,
\label{Q-mag dispersion relation}
\end{equation}
where $p$ is the momentum of the magnon bound state along the spin-chain.
The function $f(\lambda)$ should be given by $f(\lam)=\lambda/\pi^2$ in view of existing results of perturbative computations in the SYM side.
When $Q \ll \sqrt{\lambda}$, the dispersion relation (\ref{Q-mag dispersion relation}) reduces to $\Delta-J_{1}=(\sqrt{\lam}/\pi) \abs{\sin\ko{p/2}}$, which matches with the energy-spin relation for the giant magnons obtained in \cite{HM06}, 
by identifying $p$ with the angular distance between the endpoints of an ``open'' string.
The dispersion relation (\ref{Q-mag dispersion relation}) was precisely reproduced from a classical string theory computation in \cite{CDO06, AFZ06, MTT06, SV06}, where they considered a two-charge extension of giant magnon solution (``dyonic giant magnons''), identifying $Q$ of order $\sqrt{\lam}$ with the second spin $J_2$.
Thus another interesting example of the correspondence at the level solutions is obtained: $(iii)$ Dyonic giant magnons vs. Magnon bound states in an asymptotic spin-chain.

\paragraph{}
We have so far seen three interesting examples $(i)$-$(iii)$ in testing the spinning-string/spin-chain duality in the SU(2) sector at the level of concrete solutions.
They all played key r\^{o}les in checking the duality in non-BPS sectors in quite a non-trivial way.
It would be then natural to seek for more generic two-spin string solution interpolating both the BMN/Frolov-Tseytlin and the HM cases, which would give us further playground to test the AdS/CFT.
With this in mind, in the current paper, we explicitly construct a family of classical string solutions with large spins on $\mr{\mathbb R}_{\rm t}\times \mr{S}^{3}\subset \mr{AdS}_{5}\times \mr{S}^{5}$, which are related to soliton solutions of Complex sine-Gordon (CsG) theory via so-called Pohlmeyer's reduction procedure.
We will mainly focus on solutions which interpolate the spinning strings of Frolov and Tseytlin \cite{FT03} and the dyonic giant magnons. 
Our strategy of exploiting the relation between the integrable system and the string sigma model will turn out quite powerful in constructing the solutions of our concern.
Such a point of view has also worked quite efficiently in recent papers \cite{HM06, CDO06, CDO06b, Roiban06}.

We shall present the generic solutions in terms of elliptic theta functions, which should favor interpretations from the standpoint of a finite-gap problem.
We believe such a point of view will make the classification of classical string solutions more tractable.
In \cite{DV06, DV06b}, general finite-gap solutions to the equations of motions on {\RS} were constructed, and in \cite{MTT06}, finite-gap solution interpretation of giant magnons was discussed in this context.
This line of study, which originates from the work \cite{KMMZ04} and developed further toward the full sector \cite{BKSZ05}, is particularly important since it would allow us to directly compare both sides of the AdS/CFT on the level of algebraic curves.
It would be then very interesting to re-construct our solutions as finite-gap solutions.
We will be back to this point in Section \ref{sec:summary+discussion}.

\paragraph{}
This paper is organized as follows.
In Section \ref{sec:CsG soliton}, we will briefly explain Pohlmeyer's reduction procedure.
In Section \ref{sec:1-spin}, we will construct string solutions with a single large-spin in $\mr{S}^{2}$, which are related to periodic soliton solutions of sine-Gordon (sG) equation.
These strings will be classified into two types by their profiles in target space.
Then we generalize the analysis to two-spin cases in Section \ref{sec:2-spin}, in which case the strings are related to periodic soliton solutions of CsG equations.
In Section \ref{sec:limits}, we will take various limits for our solutions, and see how they interpolate various known string configurations.
Section \ref{sec:summary+discussion} will be devoted to conclusions and discussions. 
In Appendix \ref{app:Elliptic Functions}, we will present our conventions for elliptic integrals and elliptic functions.
Some formulae useful in deriving our results will be collected in Appendix \ref{app:Details of Calculations}.

\section{Classical Strings as Complex Sine-Gordon Solitons} \label{sec:CsG soliton}

In this section, we will briefly sketch how classical strings on \RS\ are related to the solitons of CsG equations.
Throughout the paper, we are concerned with classical strings moving on \RS, which is a subspace of ${\rm AdS}_{5}\times {\rm S}^{5}$ background of type IIB string theory.
Let us write the metric on \RS\ as
\begin{equation}
ds^{2}_{\mathbb R\times {\rm S}^3} = -d\eta_{0}^{2}+\abs{d\xi_{1}}^{2}+\abs{d\xi_{2}}^{2}\,,
\end{equation}
where $\eta_{0}$ is the AdS-time, and the complex coordinates $\xi_{j}$ $(j=1,2)$ are defined by the embedding coordinates $X_{M=1,\dots,4}$ of ${\rm S}^{3} \subset \bb{R}^4$ as
\begin{equation}
\xi_{1} = X_{1}+i X_{2} = \cos \theta \; e^{i \varphi_1}  \qquad {\rm and} \qquad \xi_{2} = X_{3}+i X_{4} = \sin \theta \; e^{i \varphi_2} \,.
\label{embed_coords}
\end{equation}
Here we set the radius of ${\rm S}^3$ to unity so that $\sum_{M=1}^{4} X_{M}^{2}=\sum_{j=1}^{2}\abs{\xi_{j}}^{2}=1$.
The Polyakov action for a string which stays at the center of the AdS${}_{5}$ and rotating on the three-sphere then takes the form,
\begin{equation}
S_{\ssp \bb{R}_{\rm t} \times {\rm S}^3} = - \f{\sqrt{\lambda}}{2}\int d\tau\int\f{d\sigma}{2\pi}\kkko{
\gamma^{ab}\kko{- \, \pa_{a}\eta_{\ssp 0} \, \pa_{b}\eta_{\ssp 0}+\pa_{a}\vec\xi\cdot\pa_{b}\vec\xi^{*}} + \Lambda(|\vec\xi|^{2}-1)
}\,,   \label{RtS3_action}
\end{equation}
where we used the AdS/CFT relation $\al' = 1/ \sqrt{\lam}$, and $\Lambda$ is a Lagrange multiplier.
We take the standard conformal gauge, $\gamma^{\tau \tau} = -1$, $\gamma^{\sigma \sigma} = 1$ and $\gamma^{\sigma \tau} = \gamma^{\tau \sigma} = 0$.
Let us denote the energy-momentum tensor which follows from the action (\ref{RtS3_action}) as $\cT_{ab}$,
then the Virasoro constraints are imposed as
\begin{equation}\label{string_Virasoro}
\begin{matrix}
&\quad 0 &= \ds \cT_{\sig\sig}=\cT_{\tau\tau} = - \frac12 \, (\pa_{\tau}\eta_{0})^{2} - \frac12 \, (\pa_{\sigma}\eta_{0})^{2} + \frac12 \, |\pa_{\tau}\vec\xi|^{2} + \frac12 \, |\pa_{\sig}\vec\xi|^{2} \\[3mm]
\mbox{and} &\quad 0&=\cT_{\tau\sig}=\cT_{\sig\tau}= \Re \pare{ \pa_{\tau}\vec\xi\cdot \pa_{\sig}\vec\xi^{*} } . \hfill
\end{matrix}
\end{equation}
The equations of motion that follow from \eqref{RtS3_action} are given by
\begin{equation}
\pa_{a}\pa^{a}\eta_{0}=0\quad 
\mbox{and}\quad 
\pa_{a}\pa^{\ssp a}\vec\xi+(\pa_{a}\vec\xi\cdot\pa^{\ssp a}\vec\xi^{*})\vec\xi = \vec 0\,.  \label{string_eom}
\end{equation}

\paragraph{}
Now we are going to solve the equations \eqref{string_Virasoro} and \eqref{string_eom} to find consistent string motions. 
Our strategy for that purpose is to make use of the trick invented by Pohlmeyer, that is, to relate O(4) nonlinear sigma model with conformal gauge to CsG system \cite{Pohlmeyer76}.\footnote{\,This relation was also found by Lund and Regge \cite{LR76}.}
With a solution of the CsG equations at hand, the problem of constructing corresponding string solutions will boil down to just solving a Schr\"{o}dinger equation with a potential resulted from the CsG solution.

The recipe for Pohlmeyer's reduction for O(4) sigma model is as follows. 
First, define worldsheet light-cone coordinates $\sigma^{\pm}$ by
$\tau = \sigma^+ + \sigma^-,\ \sigma = \sigma^+ - \sigma^-$.
Second, choose a basis of O(4)-covariant vectors as $X_i$, $\partial_+ X_i$, $\partial_- X_i$ and $\epsilon_{ijkl} X^j \partial_+ X^k \partial_- X^l \equiv K_i$ $(i, j, k, l = 1, \dots, 4)$ so that any vectors can be written as a linear combination of them.
We can then define two O(4)-invariants $\phi$ and $\chi$ through the relations
\begin{gather}
- \partial _+ \vec X \cdot \partial_- \vec X \equiv \cos \phi \,, 
\label{def_cos_phi} \\[2mm]
\partial _ + ^2 \vec X \cdot \vec K \equiv 2 \, \partial _ +  \chi \, \sin ^2 (\phi/2),\qquad
\partial _ - ^2 \vec X \cdot \vec K \equiv  - 2 \, \partial _ -  \chi \, \sin ^2 (\phi/2) .
\end{gather}
Third, by using the equations of motion, Virasoro constraints and the normalization condition $|\vec \xi|^{2}= 1$, write the equations of motion for $\phi$ and $\chi$ as
\begin{equation}
\partial _a \partial ^{\ssp a} \phi  - \sin \phi  - \frac{{\sin \left( {\phi /2} \right)}}{{2\cos ^3 \left( {\phi /2} \right)}}\left( {\partial _a \chi } \right)^2  = 0 \,,\quad
\partial _a \partial ^{\ssp a} \chi  + \frac{{2 \, \partial _a \phi \, \partial ^{\ssp a} \chi }}{{\sin \phi }} = 0 \,.
\label{CsG eq}
\end{equation}
They are nothing but the CsG equations.
Finally, substitute \eqref{def_cos_phi} into \eqref{string_eom} to get
\begin{equation}
\pa_{a} \pa^{\ssp a} \vec \xi + (\cos \phi) \vec \xi = \vec 0 \,.
\label{phi<->del X del X}
\end{equation}
This is the Schr\"{o}dinger equation with a self-consistent potential mentioned above.

In \cite{CDO06}, the authors utilized Pohlmeyer's reduction to obtain a family of classical string solutions called dyonic giant magnons, which were associated with {\it kink} solitons of CsG equations.
In the same spirit, we are now going to exploit so-called {\it helical wave} solutions of CsG equations to find new, more general motions of strings on {\RS}.

Before doing so, let us end this section by making some additional notes on CsG system.
The CsG equations \eqref{CsG eq} follow from the Lagrangian
\begin{equation}
{\cal L}_{\rm CsG} = \frac{1}{2}\left( {\partial _a \phi } \right)^2  + \frac{\tan ^2 (\phi / 2)}{2} \left( {\partial _a \chi } \right)^2  - \cos \phi \,.
\end{equation}
By introducing a complex field $\psi \equiv \sin (\phi / 2) \exp (i \chi / 2)$, we can rewrite it as
\begin{equation}
\cL_{\rm CsG}=\f{\tilde \pa_{a}\psi^{*}\, \tilde \pa^{a}\psi}{1-|\psi|^{2}} + \mu^{-2} |\psi|^{2}\,.
\label{CsG Lagrangian}
\end{equation}
where we have also introduced a real parameter $\mu$ to rescale the worldsheet variables as $(\tilde \tau, \tilde \sigma)\equiv (\mu\tau,\mu\sigma)$.
Then the equations of motion can be combined into
\begin{equation}
\tilde \partial _a \tilde \partial^{\ssp a} \psi  + \psi^{*} \, \frac{( \tilde \partial _a \psi )^2 }{1 - \left| \psi  \right|^2 } - \mu^{-2} \psi \left( {1 - \left| \psi  \right|^2 } \right) = 0\,.
\label{csg_eq}
\end{equation}
When $\chi={\rm constant}$, this CsG system reduces to sG system with the sG field $\phi$.

\section{Helical String Solutions with a Single Spin\label{sec:1-spin}}

To illustrate our strategy to find general classical string solutions, let us begin with a simple single-spin case.
It should result from a so-called ``helical wave'' (or ``kink train'') of sG theory, which is a rigid array of kinks.
An example of such helical solitons is given by
\begin{equation}
\phi_{\rm cn} (x,t)=2\arcsin\kko{\CN{\f{(x-x_0) - v(t-t_0)}{k\sqrt{1-v^{2}}}}{k}}\,,
\label{helical cn_wave}
\end{equation}
where $v$ is the soliton velocity, $(t_{0},x_{0})$ are initial values for $(t,x)$ which will be set to zero in what follows, and cn is the Jacobian cn function.\footnote{\,For our conventions of elliptic functions and elliptic integrals, see Appendix \ref{app:Elliptic Functions}.}
The parameter $k$ determines the spatial period (or ``wavelength'') of $\phi$ field with respect to $x-vt$ as $4 \ssp k\,\eK(k) \sqrt{1-v^{2}}$.
Note that in the limit $k\to 1$, (\ref{helical cn_wave}) reduces to an ordinary single-kink soliton with velocity $v$,
\begin{equation}
\phi(x,t)=2\arcsin\kko{{1 \mathord{\left/ {\vphantom {\cosh\ko{\f{x - vt}{\sqrt{1-v^{2}}}}}} \right. \kern-\nulldelimiterspace} \cosh\ko{\f{x - vt}{\sqrt{1-v^{2}}}}}}\,.
\end{equation}
As discussed before, our strategy to find periodic string solutions is to substitute (\ref{helical cn_wave}) into (\ref{phi<->del X del X}) to obtain a Schr\"{o}dinger equation.
For a generic helical soliton, the string equation of motion (\ref{phi<->del X del X}) can be written in the form
\begin{equation}
\bpare{- \partial_\tau^2 + \pa_{\sigma}^{2} - \mu^2 k^{2}\cpare{2\sn^{2} \pare{\frac{\mu (\sigma - v \tau)}{\sqrt{1 - v^2}},k} - 1}} \vec \xi = \mu^2 U \, \vec \xi \,,  \label{lame_0}
\end{equation}
with $(k \mu \ssp \tau, k \mu \ssp \sigma) \equiv (t,x)$.
In particular, we have $U=0$ for the cn-type helical soliton \eqref{helical cn_wave}, but we will keep $U$ general for the moment.
Let us introduce boosted worldsheet coordinates,
\begin{equation}
T(\tau,\sig) \equiv \frac{\tilde \tau - v \tilde \sigma}{\sqrt{1 - v^2}} \,,\qquad
X(\tau,\sig) \equiv \frac{\tilde \sigma - v \tilde \tau}{\sqrt{1 - v^2}} \,,
\label{def:X,T}
\end{equation}
with which we can rewrite the string equation of motion \eqref{lame_0} as 
\begin{equation}
\kko{-\pa_{T}^{2}+\pa_{X}^{2} - k^{2}\ko{2\sn^{2}(X,k)-1}} \vec\xi = \ssp U \,\vec\xi\,.
\label{reduced_eom}
\end{equation}
We can solve this equation under an Ansatz
\begin{equation}
\xi_{j} (T,X;w_{j})=  {\mathcal Y}_{j}(X;w_{j}) \, e^{i u_{j}(w_{j}) T}\qquad (j=1,2)\,.
\label{ansatz xi}
\end{equation}
Here $w_{j}$ are complex parameters and ${\mathcal Y}_{j}$ are independent of $T$. 
As for constraints on $w$, see Appendix \ref{app:Details of Calculations}.
The differential equation satisfied by ${\mathcal Y}_{j}$ then takes the form
\begin{equation}
\cpare{ \,\frac{d^2}{d X^2} - k^{2}\Big( 2 \sn^{2} \pare{X,k} - 1 \Big) +u_{j}^{2}}{\mathcal Y}_{j} = U \, {\mathcal Y}_{j} \,,  \label{Lame}
\end{equation}
which is known as {\it Lam\'{e} equation}.
General eigenfunctions of Lam\'{e} equations were found by Hermite and Halphen in the nineteenth century; see Chapter 23.7 of \cite{WW58} for details.
They are given by
\begin{equation}
{\mathcal Y}(X; w)\,  \propto\,\f{\Th_1 (X-w,k)}{\Th_{0}(X,k)}\,\exp\ko{Z_{0}(w,k) X}
\quad 
\mbox{with}\quad 
u^2 = \dn^2 (w, k) + U \,,
\label{Lame_eig_fn}
\end{equation}
where $\Theta_\nu$, $Z_\nu$ are the Jacobian theta and zeta functions defined in Appendix \ref{app:Elliptic Functions}, respectively.

The result \eqref{Lame_eig_fn} is a good starting point for us to construct string solutions that satisfy the string equation of motion (\ref{lame_0}), the consistency condition for Pohlmeyer's reduction (\ref{def_cos_phi}) and the Virasoro conditions \eqref{string_Virasoro}.
Actually it turns out that, corresponding to several possibilities of choosing a helical soliton solution of (C)sG equation, there can be as many consistent string solutions. As it seems likely that all of them are related by appropriate reparametrization of the elliptic functions, in this paper, we are only concerned with cn-type helical soliton of (\ref{helical cn_wave}).

Recall that in Gubser-Klebanov-Polyakov (GKP) case \cite{GKP02}, there were two possible configurations of closed strings moving on S${}^2$\,: the folded and circular string.
We will see, in our helical case also, there are two types of {\it rigid}\, string configurations possible.
They will turn out to reduce, in certain limits, to each of two GKP configurations. 
The first type stays only one of the hemispheres about the equator, say the northern hemisphere (See Figure \ref{fig:type1} below), while the second type sweeps in both hemispheres, crossing the equator several times (Figure \ref{fig:type2}).
We will call the first type ``type (i)'' and the second ``type (ii)'' {\it helical}\, string solution, after the name ``helical wave'' in soliton theory.
Below we will demonstrate these two types in turn.
We will only present the results, and the details will be presented in Section \ref{sec:2-spin} and Appendix \ref{app:Details of Calculations}.

\subsection{Type (i) Helical Strings with a Single Spin\label{subsec:type (i) 2-spin}}

We begin with the type (i) case.
The profile is given by
\begin{align}
\eta_{0} (T,X)&=a T+b X\qquad 
\mbox{with}\quad a =k\cn(\iom)\,,\quad b =-ik\sn(\iom)\,,
\label{eta0-f}\\
\xi_{1} (T,X)&=\f{\sqrt{k}}{\dn(\iom)}\f{\Th_{0}(0)}{\Th_{0}(\iom)}\f{\Th_{1}(X-\iom)}{\Th_{0}(X)}\, 
\exp\kko{Z_{0}(\iom)X+i\dn(\iom)T}\,,
\label{xi1-f}\\
\xi_{2} (T,X)&=\f{\dn (X)}{\dn(\iom)}\,,
\label{xi2-f}
\end{align}
with $\omega$ a real parameter.
The soliton velocity $v$, which appeared in the definitions of $T$ and $X$ \eqref{def:X,T}, is related to the parameters $a$ and $b$ in \eqref{eta0-f} as $v\eq b/a$.
Using various properties and identities listed in Appendices \ref{app:Elliptic Functions} and \ref{app:Details of Calculations}, one can check the proposed set of solutions (\ref{eta0-f})-(\ref{xi2-f}) indeed satisfies the required physical constraints.
Note that the AdS-time variable $\eta_{0}$ can be rewritten as $\eta_{0} = k \tilde \tau$.

\begin{figure}[htb]
\begin{center}
\vspace{.7cm}
\hspace{-.0cm}\includegraphics[scale=0.6]{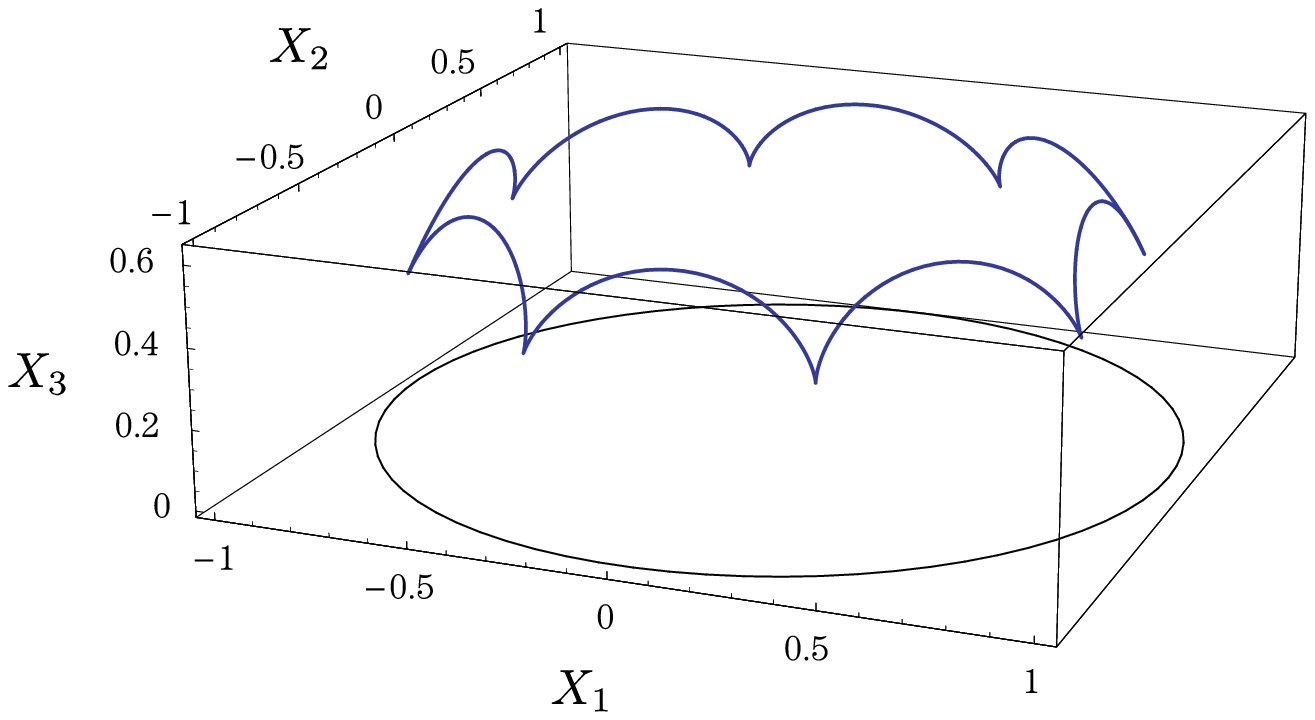}
\hspace{2.0cm}\includegraphics[scale=0.6]{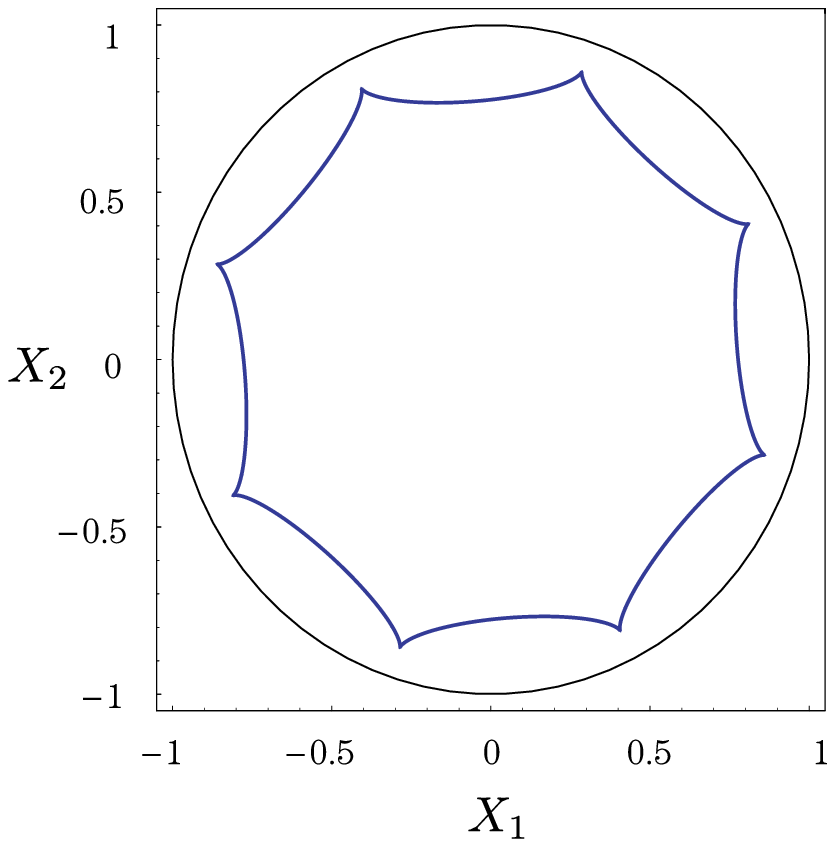}
\vspace{.5cm}
\caption{\small Type (i) helical solution with a single spin.
The diagram shows $k=0.68$ and $n=8$ case.
Each turning points are located away from the equator, and each segment curves {\it inwards}.}
\label{fig:type1}
\end{center}
\end{figure}

The spacetime profile of this kind of solutions is depicted in Figure \ref{fig:type1}.
From its appearance, it looks quite similar to the one obtained in \cite{Ryang05}, which is known as a ``spiky'' string on S${}^2$.
However, it turns out the type (i) single spin solution differs from the spiky string, in that it does not actually have singularities at the apparent spikes, as can be seen from $\left.\partial_\sigma \vec \xi\,\right|_{\sig = \pm l} = \vec 0$ with $l$ defined in (\ref{one_hop}) below.
In recent papers, the authors of \cite{KRT06} argued both the ``spiky'' strings and giant magnons can be obtained from a generalized Neumann-Rosochatius Ansatz on a string sigma model. 
In \cite{AFZ06}, in discussing the finite-spin effect on giant magnons, the authors considered a similar solution with spikes, this time with an ``open'' profile.
In our opinion, those solutions are again different from ours because they result from different Ans\"{a}tze.

In order to make the string closed and rigid, we impose a periodic boundary condition.
Since our solutions are quasi-periodic in $X$ with the period $2\,\eK$, we shall refer to the region
\begin{equation}
-l \le \sigma \le l\,,\qquad 
l \equiv \frac{\eK \sqrt{1 - v^2}}{\mu} \,,
\label{one_hop}
\end{equation}
at fixed $\tau$ as ``one-hop''. 
Then the closedness of the string requires
\begin{alignat}{2}
\Delta \sigma \Big|_{\rm one\mbox{\tiny\,-\,}hop} &\equiv \frac{2 \pi}{n} = \frac{2 \eK \sqrt{1 - v^2}}{\mu} \,,
\label{Dsigma_cl} \\[2mm]
\Delta \varphi_1 \Big|_{\rm one\mbox{\tiny\,-\,}hop} &\equiv \frac{2 \pi N_{1}}{n} = 2 \eK \pare{ - i Z_0 (\iom) + \frac{i \sn (\iom) \dn (\iom)}{\cn (\iom)} } + \pare{2 \ssp n'_1 + 1}\pi ,
\label{Dphi1_cl}
\end{alignat}
with $n = 1, 2, \ldots$\,, and $N_1\,, n'_1$ being integers.

Here $\varphi_{1}$ is the azimuthal angle defined in \eqref{embed_coords}.
When $\sigma$ runs from $0$ to $2\pi$, an array of $n$ hops winds $N_{1}$ times in $\varphi_{1}$-direction in the target space, thus making the string closed.

Let us compute the conserved charges for the type (i) strings. 
As usual, the energy $E$ and the spin $J_{1}$ are defined as
\begin{equation}
E \equiv \frac{\sqrt{\lambda}}{\pi} \, {\cal E} = \frac{n\sqrt{\lambda}}{2 \pi} \int_{-l}^{\ssp l} d \sigma \,
\partial_{\tau} \eta_{0} \,,\qquad
J_{1} \equiv \frac{\sqrt{\lambda}}{\pi} \, {\cal J}_{1} = \frac{n\sqrt{\lambda}}{2 \pi} \int_{-l}^{\ssp l} d \sigma
\Im\!\pare{\xi_{1}^{*} \partial_{\tau}\xi_{1}} \,.
\end{equation}
Then the conserved charges for this type (i) solution are computed as
\begin{equation}
{\cal E} = \frac{n k \, \eK}{\cn (\iom)} \,, \qquad
{\cal J}_{1} = \frac{n(\eK - \eE)}{\dn (\iom)} \,.
\label{1ch-f}
\end{equation}

In what follows, we will see two distinct limits that reduce the solution to two simple known examples\,; 
one is the folded string of GKP, and the other is the giant magnon of HM.

\begin{figure}[htb]
\begin{center}
\vspace{.7cm}
\hspace{-.0cm}\includegraphics[scale=0.5]{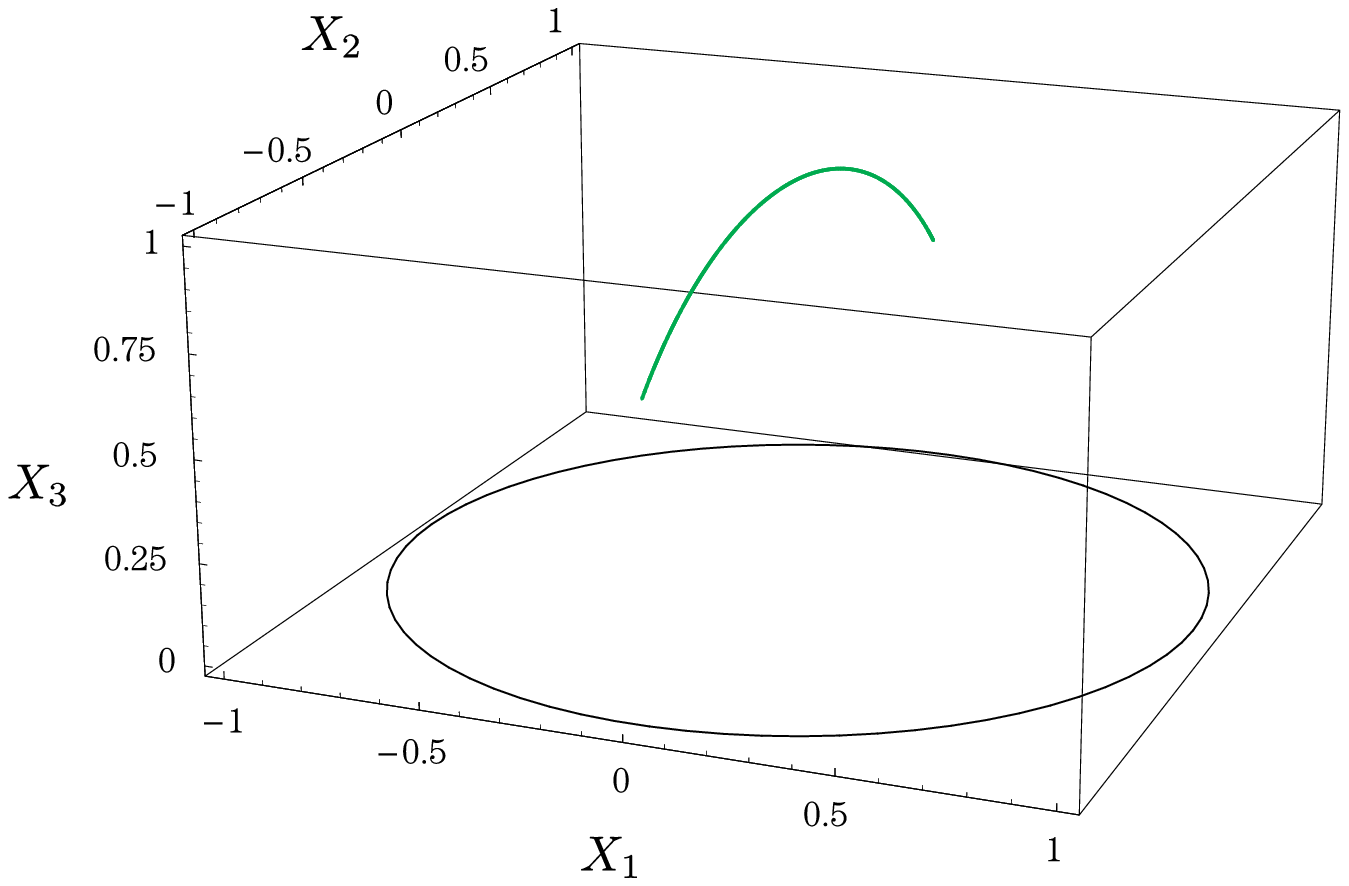}
\hspace{2.0cm}\includegraphics[scale=0.6]{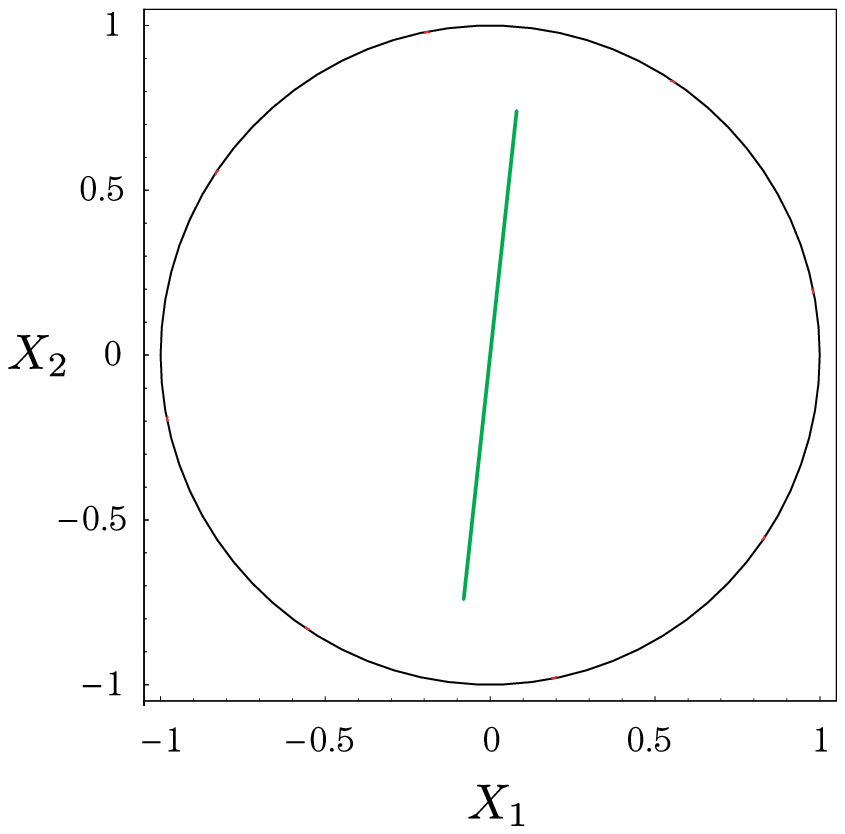}
\vspace{.5cm}
\caption{\small Type (i) helical solution with a single spin\ssp; $\om=0$ and $k=0.75$.
This can be regarded as a folded string of \cite{FT03}, in which case $n$ represents the number of folds.}
\label{fig:type1[om=0]}
\end{center}
\end{figure}

\paragraph{The GKP Case.} 
In $\om\to 0$ limit, a type (i) solution reduces to a folded string solution studied in \cite{GKP02}.
See Figure \ref{fig:type1[om=0]} for the spacetime profile.
In this limit, boosted worldsheet coordinates become 
$(T, X) \to (\tilde \tau, \tilde \sigma)$ defined in \eqref{def:X,T},
and the fields (\ref{eta0-f})-(\ref{xi2-f}) reduce to, respectively,
\begin{align}
\eta_{0}\to k \tilde \tau\,,\qquad 
\xi_{1}\to k \SN{\tilde \sig}{k}\, e^{i \tilde\tau}\,,\qquad 
\xi_{2}\to \DN{\tilde \sig}{k}\,.
\end{align}
This solution corresponds to a kink-array of sG equation {\it at rest} ($v=0$), and it spins around the northern pole of an $\mr{S}^2$ with its center of mass fixed at the pole.
The integer $n$ counts the number of folding, which is related to $\mu$ via the boundary condition (\ref{Dsigma_cl}).

\begin{figure}[htb]
\begin{center}
\vspace{.7cm}
\hspace{-.0cm}\includegraphics[scale=0.6]{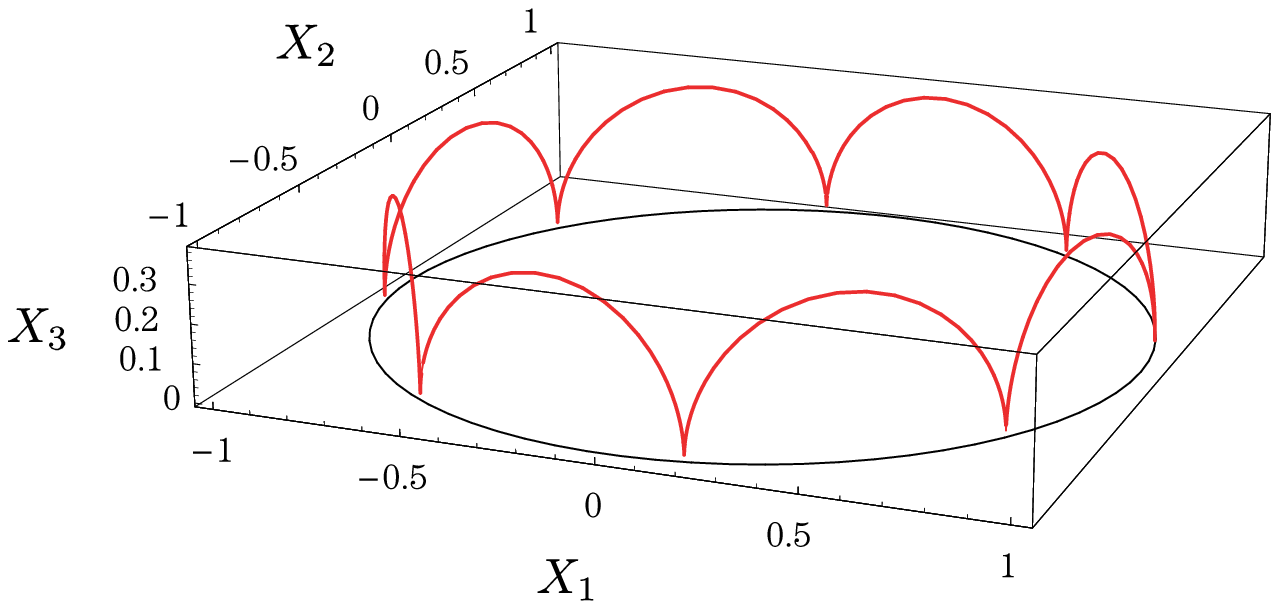}
\hspace{1.0cm}\includegraphics[scale=0.6]{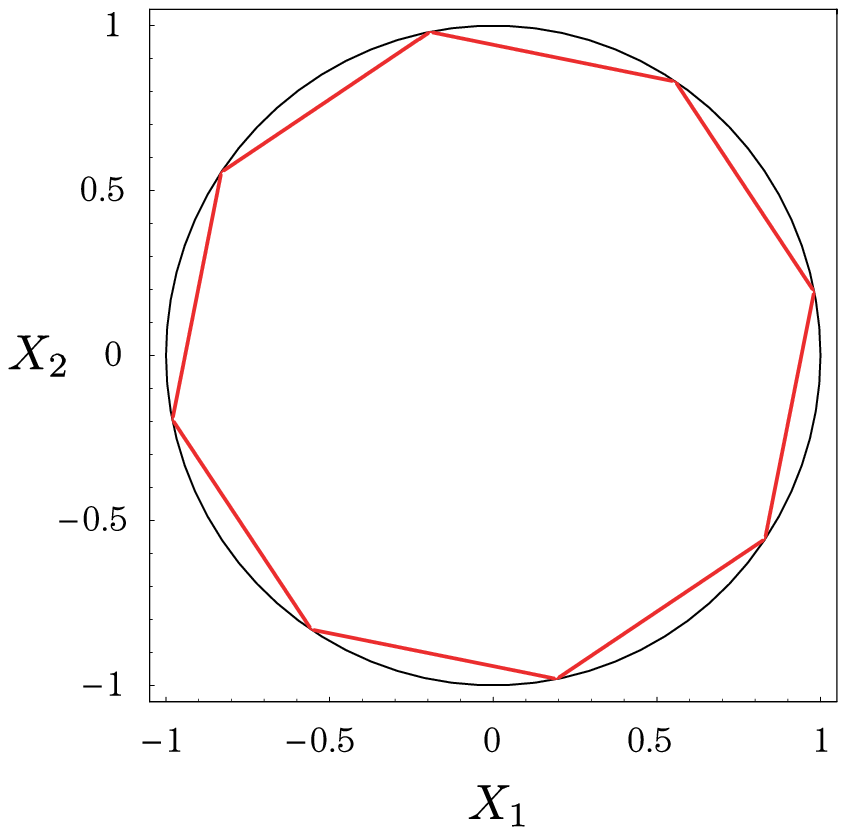}
\vspace{.5cm}
\caption{\small Type (i) helical solution with a single spin, in the limit $k\to 1$.
The diagram shows $n=8$ case, and it can be understood as an array of $n=8$ giant magnons.}
\label{fig:type1[k=1]}
\end{center}
\end{figure}

\paragraph{The HM Case.} 
The limit $k\to 1,\ \mu \to \infty$ takes the type (i) solution to an array of giant magnons, each of which having the same soliton velocity of sG system \cite{HM06}.
The endpoints of the string move on the equator $\theta=\pi/2$ at the speed of light, see Figure \ref{fig:type1[k=1]}.
In this limit, boosted worldsheet coordinates become $T\to \tilde \tau/\cos\om-\ko{\tan\om}\tilde \sig$ and $X\to \tilde \sig/\cos\om-\ko{\tan\om} \tilde \tau$, 
and the fields (\ref{eta0-f})-(\ref{xi2-f}) reduce to
\begin{align}
\eta_{0}\to \tilde \tau\,,\quad 
\xi_{1}\to \komoji{\kko{\tanh\ko{\f{\tilde\sig-\ko{\sin\om}\tilde \tau}{\cos\om}}\cos\om-i\sin\om}}\, e^{i \tilde \tau}\,,\quad 
\xi_{2}\to \f{\cos\om}{\cosh\ko{\f{\tilde \sig-\ko{\sin\om}\tilde \tau}{\cos\om}}}\,.
\label{one_giant_magnon}
\end{align}
The following boundary conditions are imposed at each end of hops\,:
\begin{equation}
\xi_{1} \to \exp\pare{\pm i \Delta \varphi_1/2 + i \tilde \tau},\quad 
\xi_{2} \to 0 \qquad {\rm as} \quad \tilde \sigma \to \pm \infty \,,
\label{bc}
\end{equation}
in place of \eqref{Dsigma_cl} and \eqref{Dphi1_cl}. 
One can see $\Delta \varphi_1$ is determined only by $\omega$, which is further related to the magnon momentum $p$ of the gauge theory as $\Delta \varphi_1 = p = \pi - 2 \omega$ in view of the AdS/CFT \cite{HM06}.

\subsection{Type (ii) Helical Strings with a Single Spin\label{subsec:type (ii) 2-spin}}

\begin{figure}[htb]
\begin{center}
\vspace{.7cm}
\hspace{-.0cm}\includegraphics[scale=0.6]{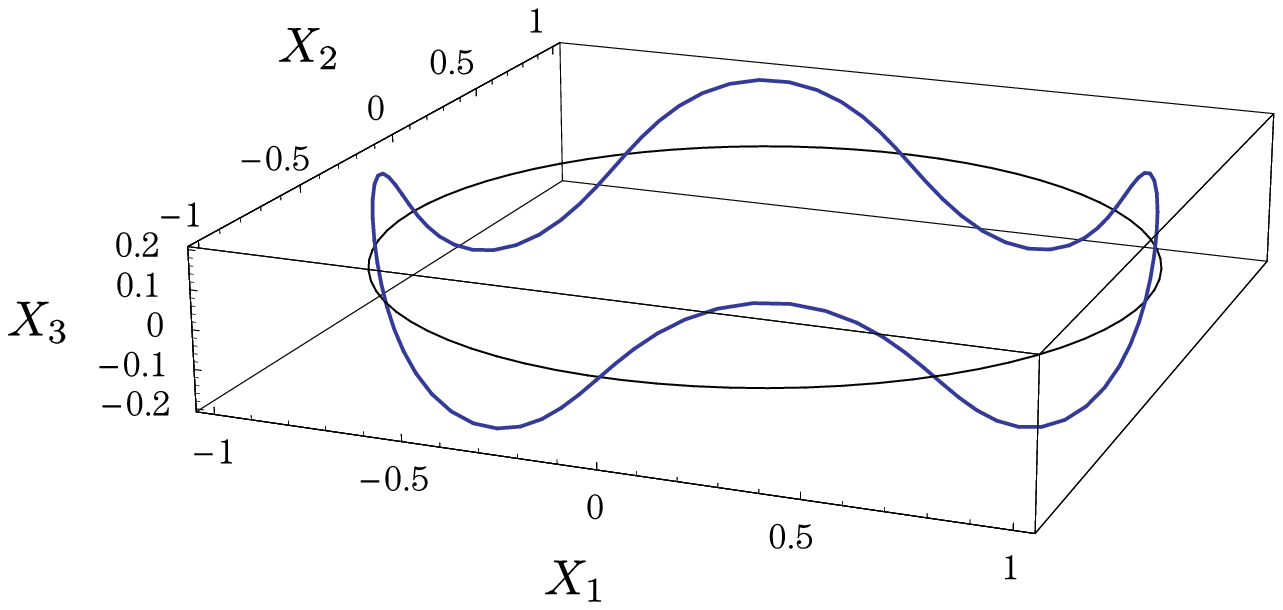}
\hspace{2.0cm}\includegraphics[scale=0.6]{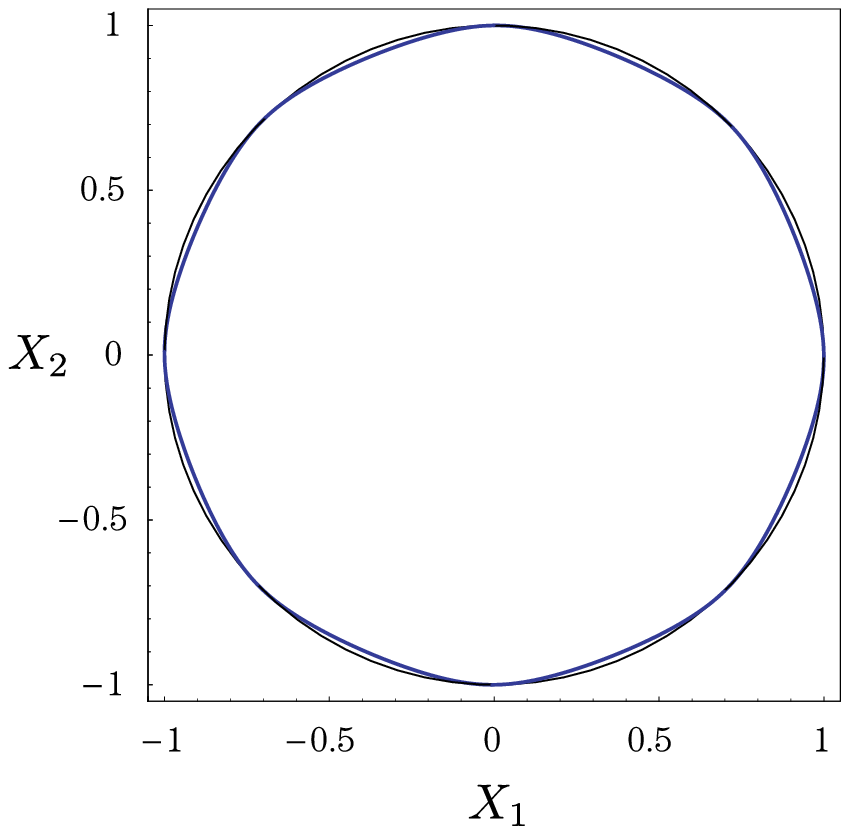}
\vspace{.5cm}
\caption{\small Type (ii) helical solution with a single spin.
The diagram  shows $k=0.68$ and $m=8$ case.
As compared to the type (i) case, each segment curves {\it outwards} about the northern pole.}
\label{fig:type2}
\end{center}
\end{figure}

Let us turn to the type (ii) solution.
In contrast to the type (i) case, it winds around the equator of $\mr{S}^{2}$, waving up and down\ssp; see Figure \ref{fig:type2}.
The profile is given by\footnote{\,We use a hat to indicate type (ii) quantities.}
\begin{align}
\hat \eta_{0} (T,X)&= \hat a \, T + \hat b X \,,\qquad 
\mbox{with}\quad \hat a =\dn(\iom)\,,\quad \hat b =-ik\sn(\iom) \,,
\label{eta0-c}\\
\hat \xi_{1} (T,X)&=\f{1}{\sqrt{k}\,\cn(\iom)}\f{\Th_{0}(0)}{\Th_{0}(\iom)}\f{\Th_{1}(X-\iom)}{\Th_{0}(X)}\, 
\exp\kko{Z_{0}(\iom)X+ik\cn(\iom)T}\,,
\label{xi1-c}\\
\hat \xi_{2} (T,X)&=\f{\cn (X)}{\cn(\iom)}\,,
\label{xi2-c}
\end{align}
where $\omega$ is again a real parameter, and the soliton velocity is given by $\hat v \equiv \hat b/\ssp \hat a$\,. 
In this type (ii) case, the AdS-time can be written as $\hat \eta_{0} = \tilde \tau$.
Just as was the case with type (i) solutions, we need to impose the following boundary conditions for a type (ii) solution to be closed\,:
\begin{alignat}{2}
\Delta \sigma \Big|_{\rm one\mbox{\tiny\,-\,}hop} &\equiv \frac{2 \pi}{m} = \frac{2 \eK \sqrt{1 - v^2}}{\mu} \,,  \label{Dsigma2_cl} \\[2mm]
\Delta \varphi_1 \Big|_{\rm one\mbox{\tiny\,-\,}hop} &\equiv \frac{2 \pi M_1}{m} = 2 \eK \pare{ - i Z_0 (\iom) + \frac{i k^2 \sn (\iom) \cn (\iom)}{\dn (\iom)} } + (2 \ssp m'_1 + 1) \pi  \,,  \label{Dphi2_cl}
\end{alignat}
where $m=1,2,\dots$ is the number of hops, $M_1$ is the winding number in $\varphi_{1}$-direction, and $m'_1$ is an integer.

The conserved charges for the type (ii) solution are calculated in the same manner as in the type (i) case.
They are given by
\begin{equation}
\hat {\cal E} = \frac{m \eK}{\dn (\iom)} \,, \qquad
\hat {\cal J} = \frac{m(\eK - \eE)}{k \cn (\iom)} \,. \label{1ch-c} 
\end{equation}

\begin{figure}[htb]
\begin{center}
\vspace{.7cm}
\hspace{-.0cm}\includegraphics[scale=0.4]{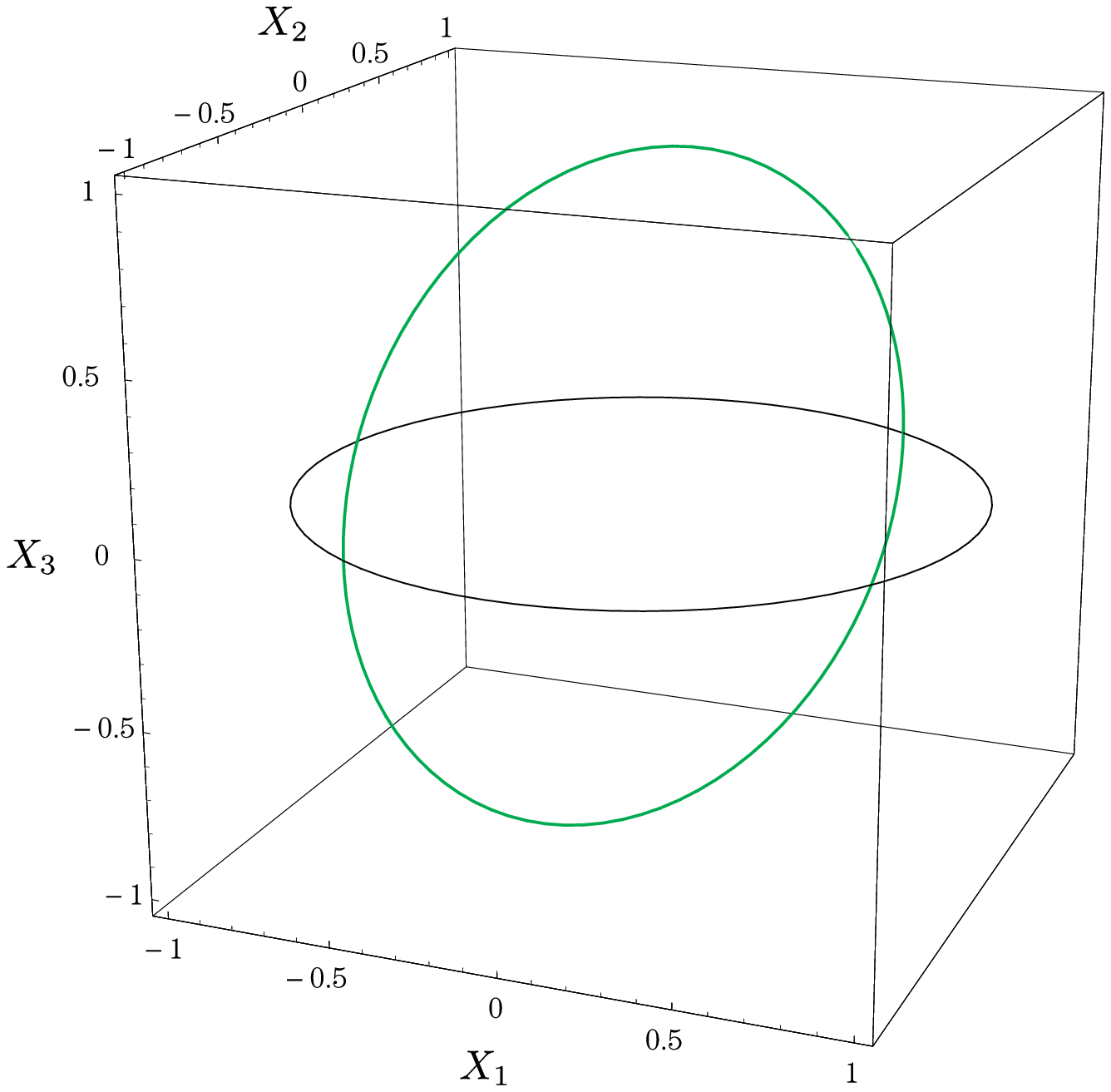}
\hspace{2.0cm}\includegraphics[scale=0.6]{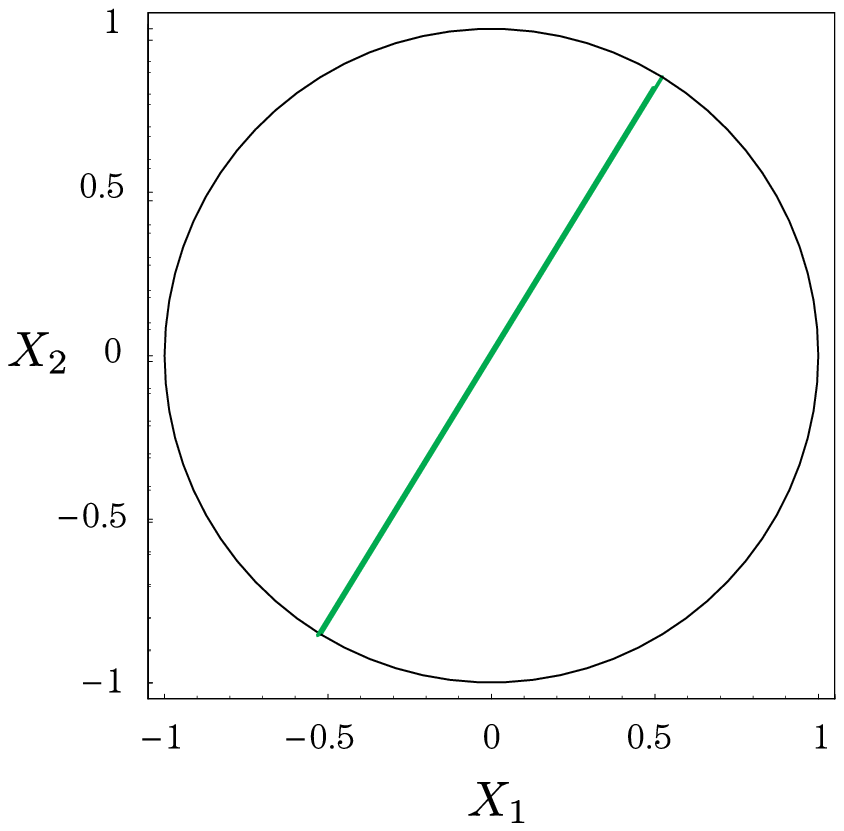}
\vspace{.5cm}
\caption{\small Type (ii) helical solution with a single spin, with $\om=0$.
This can be regarded as a circular string of \cite{FT03}, in which case $m/2$ represents the winding number along a great circle.}
\label{fig:type2[om=0]}
\end{center}
\end{figure}

\paragraph{The GKP Case.}
In $\omega \to 0$ limit, a type (ii) solutions reduce to a circular string studied in \cite{GKP02}. 
See Figure \ref{fig:type2[om=0]} for a snapshot.
Again, the boosted coordinates \eqref{def:X,T} become 
$(T, X) \to (\tilde \tau, \tilde \sigma)$, 
and the profile reduces to
\begin{equation}
\hat \eta_{0} \to \tilde \tau\,,\qquad 
\hat \xi_{1} \to \SN{\tilde \sig}{k}\, e^{i \tilde \tau}\,,\qquad 
\hat \xi_{2} \to \CN{\tilde \sig}{k}\,.
\end{equation}
The integer $m$ counts the number of winding, which is related to $\mu$ via the boundary condition (\ref{Dsigma2_cl}).

\begin{figure}[htb]
\begin{center}
\vspace{.7cm}
\hspace{-.0cm}\includegraphics[scale=0.6]{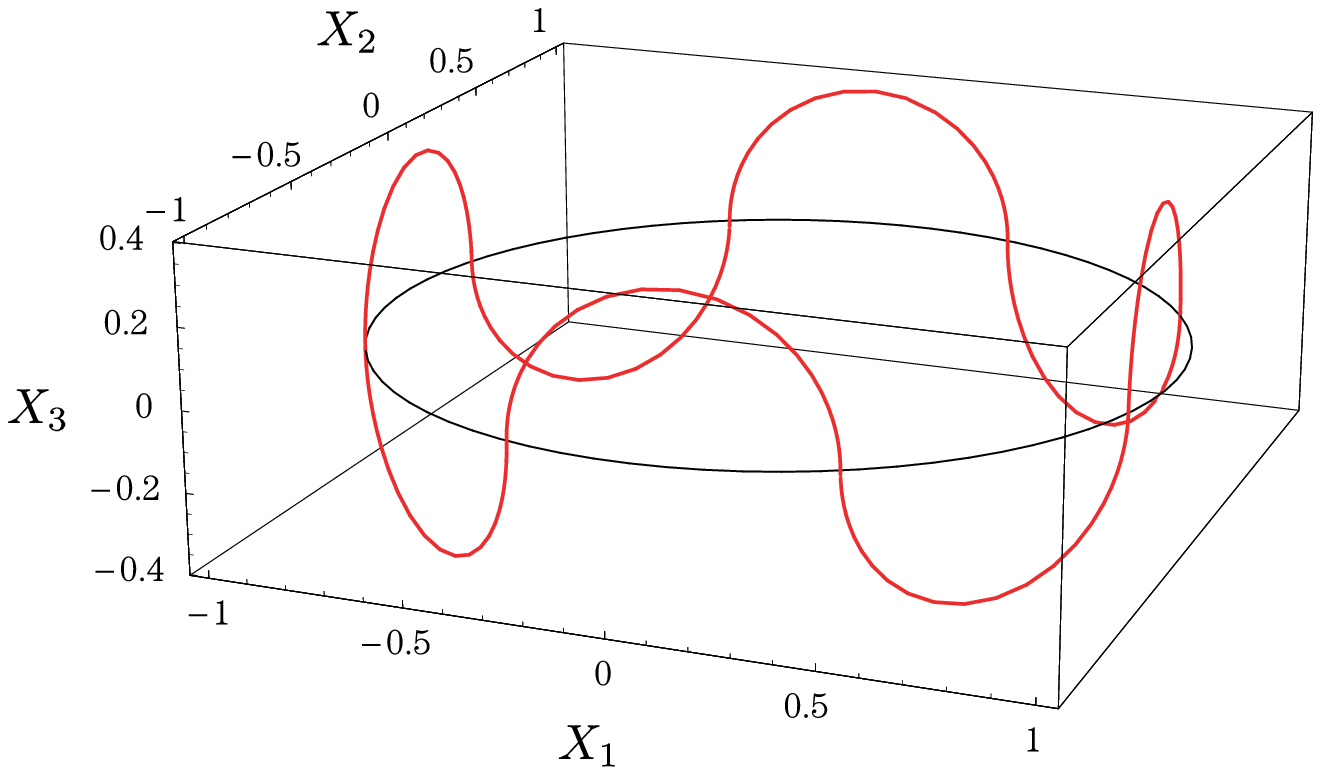}
\hspace{2.0cm}\includegraphics[scale=0.6]{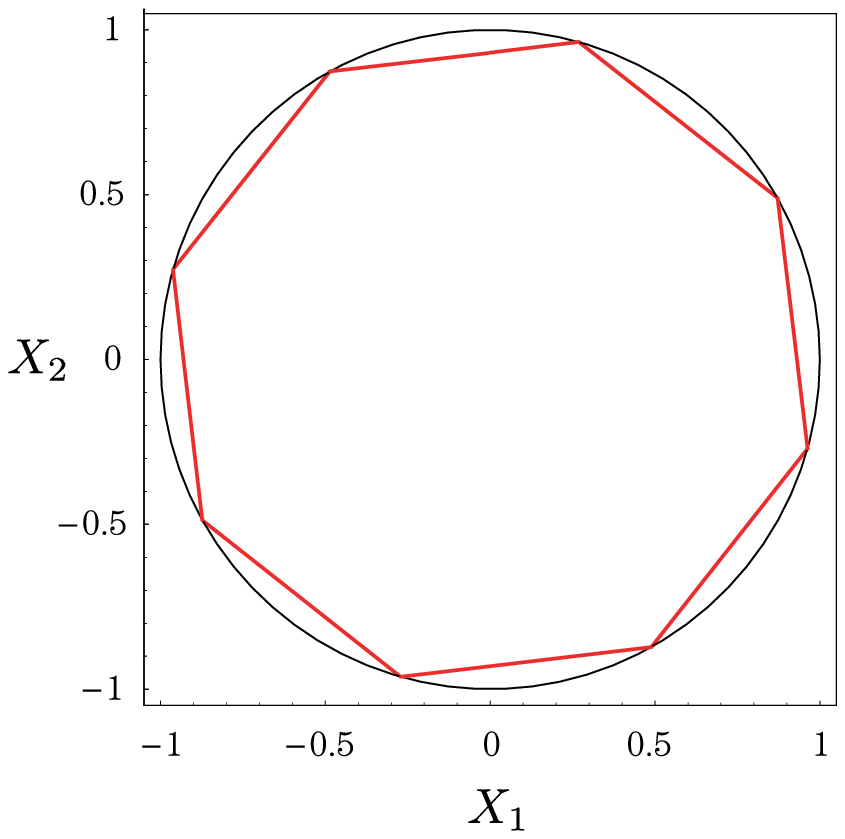}
\vspace{.5cm}
\caption{\small Type (ii) helical solution with single spin, in the limit $k\to 1$.
The diagram shows $m=8$ case, and it can be realized as an array of four
giant magnons and four flipped giant magnons by turns.
It can be regarded as the same configuration as that of Figure
\ref{fig:type1[k=1]}, which is made up of eight giant magnons; these two
configurations can be switched to each other without energy costs.}
\label{fig:type2[k=1]}
\end{center}
\end{figure}

\paragraph{The HM Case.}
The limits $k\to 1$ and $\mu \to \infty$ reduce the type (ii) solution to an
array of giant magnons and flipped giant magnons, one after the other.
The shape of each giant magnon is same as \eqref{one_giant_magnon}, see
Figure \ref{fig:type2[k=1]}.

\section{Helical Solutions with Two Spins\label{sec:2-spin}}

Let us now turn to the problem of finding generic helical string solutions with two spins.
As discussed in Section \ref{sec:CsG soliton}, string solutions on {\RS} of our concern are related to CsG solitons via Pohlmeyer's reduction. 
Therefore we begin with generalizing helical solitons of sG equation \eqref{helical cn_wave} to those of CsG equations.
One can easily confirm the following function is an example of such helical solutions of CsG equations:
\begin{equation}
\psi_{\rm cn} = c \ssp k \, \cn\left( c \ssp x_v ,k \right)\exp \left( {i \ssp t_v \sqrt {(1 - c^2 k^2)(1 + c^2 (1 - k^2))} } \right) ,  
\label{helical_cn}
\end{equation}
where $c$ takes the value in $-1/k<c<1/k$ for $0 \le k \le 1$, and $x_v$, $t_v$ are defined as
\begin{equation}
x_v \equiv \frac{x - v t}{\sqrt{1 - v^2}} \,,\qquad
t_v \equiv \frac{t - v x}{\sqrt{1 - v^2}} \,.
\label{boosted_co}
\end{equation}
Thus the periodic function \eqref{helical_cn} can be thought of a natural generalization of (\ref{helical cn_wave}).
We will use this solution to find the dyonic extended version of helical solutions.

The string equations of motion become the same as \eqref{reduced_eom} under identifications $(\mu \tau, \mu \sigma) \equiv (c \ssp t, c \ssp x)$, and we can solve them with the same Ansatz (\ref{ansatz xi}).
For the case of cn-type helical soliton \eqref{helical_cn}, $U$ is evaluated as $U_{\rm cn} = (1/c^2) - k^2 \ge 0$.
If we started with other helical solitons such as of sn- or dn-type, they would give different ranges for $U$ in general.
Hence we will treat $U$ as a controllable parameter.

We are interested in string configurations with two spins, which interpolate known string solutions in an obvious way.

\subsection{Type (i) Helical Strings with Two Spins}
First we will focus on the type (i) case.
The solution can be written in the following form\ssp:
\begin{align}
\eta_{0} &= a T + b X\,,   \label{zf0}  \\[2mm]
\xi_{1} &= C \frac{\Theta _0 (0)}{\sqrt{k} \, \Theta _0 (\iomm{1})} \frac{\Theta _1 (X   - \iomm{1})}{\Theta _0 (X )} \,
\exp \Big(  Z_0 (\iomm{1})  X+ i  u_1 T\Big)\,,
\label{zf1} \\[2mm]
\xi_{2} &= C \frac{\Theta _0 (0)}{\sqrt{k} \,\Theta _2 (\iomm{2})} \frac{\Theta _3 (X   - \iomm{2})}{\Theta _0 (X )} \, \exp \Big( Z_2 (\iomm{2} )X + i   u_2 T\Big)\,.
\label{zf2}
\end{align}
Here $\omega_1$ and $\omega_2$ are real parameters.
The normalization constant $C$ is chosen as
\begin{equation}
C = \pare{ \frac{\dn^2 (\iomm{2})}{k^2 \cn^2 (\iomm{2})} - 
\sn^2 (\iomm{1}) }^{-1/2} ,
\end{equation}
so that the sigma model condition $\left| \xi_{1} \right|^2 + \left| \xi_{2} \right|^2 = 1$ is satisfied.
The parameters $a$ and $b$ in (\ref{zf0}) are fixed by Virasoro conditions, which imply
\begin{align}
a^2 + b^2 &= k^2 - 2 k^2 \sn^2 (\iomm{1}) - U + 2 \ssp u_2^2 \,,
\label{a,b-f1}  \\
\quad a \ssp b &= - i \, C ^2
\pare{u_1 \sn (\iomm{1}) \cn (\iomm{1}) \dn (\iomm{1}) - u_2 \, \frac{1-k^2}{k^2} \, \frac{\sn (\iomm{2}) \dn (\iomm{2})}{\cn^3 (\iomm{2})} }\label{a,b-f2}\,.
\end{align}
Just as in the single spin cases, we can adjust the soliton velocity $v$ so that the AdS-time is proportional to the worldsheet time variable. 
It then follows that $v \equiv b/a \le 1$ and $\eta_{0} = \sqrt{a^2 - b^2} \, \tilde\tau $.
Two angular velocities are constrained as 
\begin{equation}
u_1^2 = U + \dn^2 (\iomm{1}) \,,\qquad
u_2^2 = U - \frac{(1 - k^2) \sn^2 (\iomm{2})}{\cn^2 (\iomm{2})} \,,
\label{u_1 and u_2}
\end{equation}
where the parameter $U$ corresponds to the eigenvalue of the Lam\'{e} equation (\ref{reduced_eom}).
From \eqref{u_1 and u_2} we find the two angular velocities $u_{1}$ and $u_{2}$ satisfy
\begin{equation}
u_1^2 - u_2^2 = \dn^2 (\iomm{1}) + \frac{(1 - k^2) \sn^2 (\iomm{2})}{\cn^2 (\iomm{2})} \,.
\label{u1-u2:f}
\end{equation}
When $\omega_2 = u_2 = 0$, this reproduces the type (i) single spin solution of Section \ref{subsec:type (i) 2-spin}.
The consistency condition \eqref{def_cos_phi} is indeed satisfied as
\begin{equation}
\frac{1}{\mu^2} \sum_{i=1}^2 \pare{ \left| \partial_\sigma \xi_{i} \right|^2
- \left| \partial_\tau \xi_{i} \right|^2 } = k^2 
- 2 k^2 \sn^2 (X ) - U\,,
\end{equation}
from which we can deduce the equation of motion \eqref{reduced_eom}.

As in the single spin case, we can write down the conditions for a type (i) dyonic helical string to be closed. They read,
\begin{alignat}{2}
\Delta \sigma \Big|_{\rm one\mbox{\tiny\,-\,}hop} &\equiv \frac{2 \pi}{n} = \frac{2 \eK \sqrt{1 - v^2}}{\mu} \,,  \label{Dsigma_cl_dy} \\[2mm]
\Delta \varphi_1 \Big|_{\rm one\mbox{\tiny\,-\,}hop} &\equiv \frac{2 \pi N_1}{n} = 2 \eK \pare{ -i Z_0 (\iomm{1}) - v \ssp u_1 } + (2 \ssp n'_1 + 1) \pi \,, \label{Dphi1_cl_dy} \\[2mm]
\Delta \varphi_2 \Big|_{\rm one\mbox{\tiny\,-\,}hop} &\equiv \frac{2 \pi N_2}{n} = 2 \eK \pare{ -i Z_2 (\iomm{2}) - v \ssp u_2 } + 2 \ssp n'_2 \ssp \pi \,.  \label{Dphi2_cl_dy}
\end{alignat}
As $\sigma$ runs from $0$ to $2\pi$, the string hops $n$ times in the target space, winding $N_{1}$ and $N_{2}$ times in $\varphi_1$- and $\varphi_2$-direction, respectively.

Global conserved charges can be computed just as was done in Section \ref{sec:1-spin}.
The rescaled energy $\cE$ and the spins $\cJ_{j}$ $(j=1,2)$ are evaluated after a little algebra to give 
\begin{align}
\cE &= n \ssp a \pare{1 - v^2 }  \eK \,, \\[2mm]
\cJ_1  &= \frac{n \ssp C^2 \, u_1}{{k^2 }}\kko{ { - \eE + \left( {\dn^2 (\iomm{1}) + \frac{v \ssp k^2}{{u_1 }} \ssp i \sn(\iomm{1}) \cn(\iomm{1}) \dn (\iomm{1}) } \right)\eK} } \,, \\[2mm]
{\cal J}_2  &= \frac{n \ssp C^2 \, u_2}{{k^2 }}\cpare{ \eE + (1-k^2) \left( 
\frac{\sn^2 (\iomm{2})}{\cn^2 (\iomm{2})} - \frac{v}{u_2} \frac{i \sn(\iomm{2})\dn(\iomm{2})}{\cn^3(\iomm{2})} \right)\eK} .
\end{align}

\subsection{Type (ii) Helical Strings with Two Spins}

Next let us turn to the type (ii) solutions.
We can reach them by shifting the parameter $\omega_2$ of a type (i) solution by $\eK'$.\footnote{\,The type (ii) solution can be also obtained by applying a transformation $k \to 1/k$ to the type (i) solution, just as for the cases with the Frolov-Tseytlin solutions. See, for example, \cite{KMMZ04}.}
The resulting expressions are
\begin{align}
\hat \eta_{0} &= \hat a \ssp T + \hat b X\,,   \label{zc0}  \\[2mm]
\hat \xi_{1} &= \hat C \frac{\Theta _0 (0)}{\sqrt{k} \, \Theta _0 (\iomm{1})} \frac{\Theta _1 (X   - \iomm{1})}{\Theta _0 (X )} \,
\exp \Big( Z_0 (\iomm{1}) X + i u_1 T\Big)\,,
\label{zc1} \\[2mm]
\hat \xi_{2} &= \hat C \frac{\Theta _0 (0)}{\sqrt{k} \,\Theta _3 (\iomm{2})} \frac{\Theta _2 (X   - \iomm{2})}{\Theta _0 (X )} \, \exp \Big( Z_3 (\iomm{2} ) X+ i u_2 T \Big)\,,
\label{zc2}
\end{align}
where $\hat C$ is the normalization constant given by
\begin{equation}
\hat C = \pare{ \frac{\cn^2 (\iomm{2})}{\dn^2 (\iomm{2})} - \sn^2 (\iomm{1}) }^{-1/2}\,.
\end{equation}
The Virasoro conditions constrain the coefficients $\hat a$, $\hat b$ as
\begin{alignat}{2}
&\hat a^2 + \hat b^2 &\ &= \quad k^2 - 2 k^2 \sn^2 (\iomm{1}) - U + 2 \ssp u_2^2 \,, \label{a,b-c1}\\
&\quad \hat a \, \hat b &\ &= - i \, \hat C^2
\pare{u_1 \sn (\iomm{1})\cn(\iomm{1}) \dn(\iomm{1}) + u_2 \pare{1-k^2} \frac{\sn (\iomm{2})\cn(\iomm{2})}{\dn^3(\iomm{2})}} \,.
\label{a,b-c2}
\end{alignat}
The soliton velocity is given by $\hat v \equiv \hat b/\ssp \hat a \le 1$ so that we have 
$\hat \eta_{0} = \sqrt{\hat a^2 - \hat b^2}\,\tilde \tau$.
The angular velocities $u_{1}$ and $u_{2}$ satisfy
\begin{equation}
u_1^2 = U + \dn^2 (\iomm{1}) \,,\qquad
u_2^2 = U + \frac{1 - k^2}{\dn^2 (\iomm{2})} \,,
\end{equation}
and are constrained as
\begin{equation}
u_1^2 - u_2^2 = \dn^2 (\iomm{1}) - \frac{1 - k^2}{\dn^2 (\iomm{2})} \,.
\label{u1-u2:c}
\end{equation}
When $\omega_2 = u_2 = 0$, it reduces to the type (ii) single spin solution.

The closedness conditions for a type (ii) solution are given by
\begin{alignat}{2}
\Delta \sigma \Big|_{\rm one\mbox{\tiny\,-\,}hop} &\equiv \frac{2 \pi}{m} = \frac{2 \eK \sqrt{1 - \hat v^2}}{\mu} \,,  \label{Dsigma2_cl_dy} \\[2mm]
\Delta \varphi_1 \Big|_{\rm one\mbox{\tiny\,-\,}hop} &\equiv \frac{2 \pi M_1}{m} = 2 \eK \pare{ - i Z_0 (\iomm{1}) - \hat v \ssp u_1 } + (2 \ssp m'_1 + 1) \pi  \,,  \label{Dphi3_cl_dy} \\[2mm]
\Delta \varphi_2 \Big|_{\rm one\mbox{\tiny\,-\,}hop} &\equiv \frac{2\pi M_2}{m} = 2 \eK \pare{ - i Z_3 (\iomm{2}) - \hat v \ssp u_2 } + (2 \ssp m'_2 + 1) \pi  \,, \label{Dphi4_cl_dy}
\end{alignat}
where $m=1,2,\dots$ is again the number of hops for $0\leq \sigma \leq 2\pi$, and $M_{1}$ and $M_{2}$ are winding numbers for $\varphi_1$- and $\varphi_2$-direction, respectively.

The conserved charges of $m$ hops can be evaluated as
\begin{align}
\hat \cE &= m \ssp \hat a \pare{1 - \hat v^2 } \; \eK \,, \\[2mm]
\hat \cJ_1  &= \frac{m \ssp \hat C^2 \, u_1 }{{k^2 }}\kko{ { - \eE + \left( {\dn^2 (\iomm{1}) + \frac{\hat v \ssp k^2}{{u_1 }} \ssp i \sn (\iomm{1}) \cn (\iomm{1}) \dn (\iomm{1}) } \right)\eK} } \,, \\[2mm]
\hat \cJ_2  &= \frac{m \ssp \hat C^2 \, u_2 }{{k^2 }}\kko{ \eE - (1-k^2) \pare{ \frac{1}{\dn^2 (\iomm{2})} - \frac{\hat v \ssp k^2}{u_2} \, \frac{i \sn (\iomm{2})\cn(\iomm{2})}{\dn^3(\iomm{2})}} \eK }\,.
\end{align}

\section{Taking Various Limits\label{sec:limits}}

Now that we have obtained generic helical solutions with two spins, for both type (i) and (ii) dyonic solutions, we can reproduce known string configurations as their special limiting cases.
Interesting limits are the ``stationary'' limit $\om_{i}\to 0$, the ``infinite spin'' limit $k\to 1$ and the ``uniform charge-density'' limit $k\to  0$.
We will see them in turn.

\subsection{Stationary Limit\,:~ Frolov-Tseytlin Strings}

In the stationary limit where both $\om_{i}$ vanish, the soliton velocity tends to zero, thus reducing the solutions to 
the spinning strings of Frolov and Tseytlin \cite{FT03}.

As usual, let us begin with the type (i) case.
In this limit, the boosted coordinates \eqref{def:X,T} become $(T, X) \to (\tilde \tau, \tilde \sigma)$, 
and (\ref{zf0})-(\ref{zf2}) reduce to
\begin{equation}
\eta_{0} = \sqrt{k^2 + u_2^2} \ \tilde \tau \,, \qquad 
\xi_{1} = k \sn(\tilde \sigma, k) \, e^{i u_1 \tilde \tau} \,, \qquad 
\xi_{2} = \dn(\tilde \sigma, k) \, e^{i u_2 \tilde \tau} \,,
\end{equation}
with a constraint $u_{1}^{2}-u_{2}^{2}=1$.
This is the folded spinning/rotating string of \cite{FT03}, which stretches over a great circle in the $\theta$-direction and spinning around its center of mass with angular momentum $J_{2}$. 
The center of mass itself moves along another orthogonal great circle of {\S} with spin $J_{1}$.
To compare our results with the one presented in \cite{FT03}, one should relate the parametrization as
\begin{equation}
\tilde \tau = \mu \ssp \tau_{\ssp \rm FT} \,,\quad 
\tilde \sig = \mu \ssp \sigma_{\ssp \rm FT} \,,\quad 
\kappa_{\ssp \rm FT} = \mu \sqrt{k^2 + u_2^2}\,,\quad
w_{i} = \mu \ssp u_{i}\qquad 
\mbox{with}\quad 
\mu \equiv \sqrt{w_{1}^{2}-w_{2}^{2}}\,.
\end{equation}
In this stationary limit, 
the conserved charges take the following simple form,
\begin{align}
\cE = n\sqrt{k^2 + u_2^2} \; \eK \,, \quad 
\cJ_1 =n u_1 \pare{\eK - \eE} ,  \quad 
\cJ_2 =n u_2 \, \eE \,,
\label{charges-fold}
\end{align}
with the hopping number $n$ now represents the folding number.

By expanding the moduli $k$ and the charges $E$ and $J_{i}$ in powers of $\lam/J^{2}$ with $J\eq J_{1}+J_{2}$, we can solve (\ref{charges-fold}) order by order to obtain the energy coefficients $c_{k}$ introduced in Introduction.
They can be compared to $a_{k}$ obtained from a double-contour distribution of Bethe roots on the gauge side \cite{BMSZ03}.

\paragraph{}
Circular strings of Frolov-Tseytlin \cite{FT03} are also reproduced in much the same way, by taking the stationary limit for the type (ii) solutions.
In this case (\ref{zc0})-(\ref{zc2}) reduce to
\begin{equation}
\hat \eta_{0} = \sqrt{1 + u_2^2} \ \tilde \tau \,, \qquad
\hat \xi_{1} = \sn(\tilde \sigma, k) \, e^{i u_1 \tilde \tau} \,, \qquad
\hat \xi_{2} = \cn(\tilde \sigma, k) \, e^{i u_2 \tilde \tau} \,,
\end{equation}
with a constraint $u_{1}^{2}-u_{2}^{2}=k^{2}$.
This string wraps around a great circle of $\mr{S}^{5}$ and rotates both in $X^{1}$\ssp -\ssp$X^{2}$ and $X^{3}$\ssp -\ssp$X^{4}$ planes.
The conserved charges are given by
\begin{equation}
\cE = m\sqrt{1 + u_2^2} \; \eK \,, \quad 
\cJ_1 = \frac{m u_1}{k^2} \pare{\eK - \eE} \,, \quad 
\cJ_2 = \frac{m u_2}{k^2} \pare{\eE - (1-k^2) \eK} \,,
\end{equation}
with $m$ now represents the winding number for $\theta$-angle.

Again, the moduli $k$ and the charges can be expanded in powers of $\lam/J^{2}$ to obtain $c_{k}$.
This time, they can be compared to the $a_{k}$ for a so-called imaginary root distribution of Bethe roots on the gauge side \cite{BMSZ03}.

\subsection{Infinite Spin Limit\,:~ Dyonic Giant Magnons}

When the moduli parameter $k$ goes to unity, both type (i) and (ii) solutions become an array of dyonic giant magnons.
The relation (\ref{u1-u2:f}) (or (\ref{u1-u2:c})) implies that the $\omega_2$-dependence of the solutions disappears in this limit.
We will therefore write $\omega$ in place of $\omega_1$\ssp.
The relation $u_{1}^{2}-u_{2}^{2}=1+\tan^{2}\om$ implies $a=u_{1}$ and $b=\tan\om$ in view of (\ref{a,b-f1}) and (\ref{a,b-f2}) (or (\ref{a,b-c1}) and (\ref{a,b-c2})), and
the profiles of both types of strings become
\begin{align}
\eta_{0} = \sqrt{1 + u_2^2}\, \tilde \tau  \,,\qquad 
\xi_{1} = \frac{\sinh(X  - \iom)}{\cosh(X )}
\, e^{i \tan(\omega)X + i u_1 T}\,,\qquad 
\xi_{2} = \frac{\cos (\omega)}{\cosh(X )} \;
\, e^{i u_2 T} \,.
\end{align}
Let us impose the same boundary conditions as in the single spin case (\ref{bc}), then it requires $\mu\to \infty$ as well as the relation $\Delta \varphi_1 = \pi - 2 \omega$.

The conserved charges for one-hop ({\it i.e.}, single giant magnon) are given by
\begin{equation}
\cE = \komoji{u_1 \pare{1 - \frac{\tan^2 \omega}{u_1^2}} \eK (1)} \,, \quad
\cJ_1 = \komoji{u_1 \kko{ \pare{1 - \frac{\tan^2 \omega}{u_1^2}} \eK (1) - \cos^2 \omega }}  \,, \quad
\cJ_2 = u_2 \cos^2 \omega \,,
\label{charges DGM}
\end{equation}
where $\eK (1)$ is divergent, {\it i.e.}, $\cE$, $\cJ_1\to \infty$.
Energy-spin relation then becomes
\begin{equation}
\cE - \cJ_1 = \sqrt{ \cJ_2^2 + \cos^{2}\omega }\,.
\label{energy_spin_CDO}
\end{equation}
By comparing \eqref{energy_spin_CDO} with \eqref{Q-mag dispersion relation} with an identification $Q\eq J_{2}=(\sqrt{\lam}/\pi)\cJ_{2}$, we find $p = \pi - 2 \omega$ as we mentioned earlier. 
It would be useful to note that, one can match the expressions above with the ones presented in \cite{CDO06}, by redefining the parameters as
\begin{equation}
T = \abs{\cos \alpha} \, \widetilde T\,, \quad
X = \abs{\cos \alpha} \, \widetilde X 
\qquad {\rm and} \qquad
u_2 \equiv \tan \alpha \,,
\end{equation}
where $\widetilde T$ and $\widetilde X$ are the boosted worldsheet variables used in \cite{CDO06}.

\subsection{Uniform Charge-Density Limit}

Another interesting limit is $k \to 0$, where the densities of $J_{i}$ tend to distribute uniformly along the worldsheet space variable $\sig$ in our gauge choice.

As for the type (i) case, the parameters $a$ and $b$ go to $a \to 1$ and $b \to 0$, and the fields become
\begin{equation}
\eta_{0} = \tilde\tau\,,\qquad 
\xi_{1} =0\,,\qquad 
\xi_{2} =e^{i \tilde\tau}\,,
\end{equation}
and the conserved charges for one-hop are $E=\sqrt{\lam}/2$, $J_{1}=0$ and $J_{2}=\sqrt{\lam}/2$.
This is a point-like, BPS $(E-J_{2}=0)$ string, rotating along the great circle in the $X^{3}$\ssp -\ssp$X^{4}$ plain.

\paragraph{}
For the type (ii) case, the profile becomes
\begin{align}
\hat \eta_{0} = \sqrt{\hat a^2 - \hat b^2} \; \tilde \tau\,,\quad
\hat \xi_{1} = \hat C \; \sin (X - \iomm{1}) \, e^{i u_1 T}\,,\quad
\hat \xi_{2} = \hat C \; \cos (X - \iomm{2}) \, e^{i u_2 T}\,,
\end{align}
where $\hat C = \pare{ \cosh^2 \omega_2 + \sinh^2 \omega_1 }^{-1/2}$. The angular velocities satisfy $u_1^2 = u_2^2 = U + 1$.
The parameters $\hat a$ and $\hat b$ (with $\hat a \ge \hat b$) are determined by
\begin{alignat}{2}
&\hat a^2 + \hat b^2 &\ &= - U + 2 u_2^2  \,,\\
&\quad \hat a \, \hat b &\ &= \hat C^2 \sqrt{U + 1}
\pare{\sinh\omega_1 \cosh \omega_1 \mp \sinh\omega_2 \cosh \omega_2 } ,
\label{pm}
\end{alignat}
where $\mp$ reflects the sign ambiguity of angular momenta.
The conserved charges for one-hop are evaluated as
\begin{align}
\hat \cE &= \frac{\pi \hat a \pare{1 - \hat  v^2}}{2}  \,, \\[2mm]
\hat \cJ_1  &= - \frac{\pi \ssp \hat C^2 \hat v}{2} \, \sinh \omega_1 \cosh \omega_1 \,, \label{charges unif IIa}  \\[2mm]
\hat \cJ_2  &= \frac{\pi \ssp \hat C^2 \hat v}{2} \, \sinh \omega_2 \cosh \omega_2 \,.
\label{charges unif IIb}
\end{align}
As we are assuming $\hat a \ge \hat b \ge 0$, the situation $\hat b =0$ can be realized when $\om_{1}=\om_{2}$ with ``$-$'' sign of (\ref{pm}), or when $\om_{1} = - \om_{2}$ with ``$+$'' sign.
In both cases, the soliton velocity $\hat v \eq \hat b/ \ssp \hat a$ vanishes, which then implies the equal spin relation $J_{1}=J_{2}$ in view of 
\eqref{charges unif IIa} and \eqref{charges unif IIb}.
This equal two-spin (or ``rational'') solution can also be realized as $J_{1}=J_{2}$ case of a so-called constant-radii string solution, which follows from an Ansatz $\xi_{j}=a_{j}\, e^{i\ko{w_{j}\tau+n_{j}\sigma}}$ $(j=1,2)$ with $a_{j}$ constants \cite{FT03}.
From the viewpoint of a finite-gap problem, an equal two-spin case mentioned above corresponds to a single-cut limit of the symmetric two-cut imaginary root solution, that is, the limit when the outer two branch points of the cuts go to $\pm i\infty$, thus making it a single-cut.
This situation can also be realized as a certain limiting configuration of a single cut distribution of Bethe roots, that is, when the filling fraction of the spin-chain (the ratio of the number of impurities to the number of sites) goes to 1/2.

\section{Summary, Discussions, and Outlook\label{sec:summary+discussion}}

In this paper, we explored the SU(2) sector of a string sigma model on $\mr{AdS}_{5}\times \mr{S}^{5}$, and constructed new classical string solutions with large spins.
They corresponded to helical soliton solutions of CsG equations on a circle via Pohlmeyer's reduction procedure, and had two adjustable parameters associated with the solitons: one was the soliton velocity $v$ and the other was the moduli parameter $k$ that controlled the period of the kink-array.
The string solutions exhibited the following interesting interpolation behavior; in $k\to 1$ limit, they reduced to dyonic giant magnons, while in $v\to 0$, they became folded and circular strings of \cite{FT03}.

\paragraph{}
There can be many interesting application of our results.
We will finish this paper by discussing a few of them with some outlooks.

\paragraph{Finite-gap solutions.}

As is clear from the typical form of the profile functions, they are closely related to the Baker-Akhiezer function
\ssp; see \cite{DV06, DV06b} and references therein.\footnote{\,We would like to thank B.~Vicedo for illuminating comments concerning the interpretation of our results as finite-gap solutions.}
From their profiles, one can see our generic solutions are described by two cuts in the spectral parameter plane, just as were the well-known cases with folded and circular strings.
In order for the charges to be real, the four branch-points of the cuts must satisfy so-called reality constraint, that is, they must locate symmetrically with respect to the real axis.
Then the most general Ansatz for the location of the branch-points $\{ x_{k} \}_{k=1,\dots,4}\in {\mathbb C}$ for our helical solutions can be written as $x_{1}=i\rho\,e^{-i(\al+\delta)}$, $x_{2}=x_{1}^{*}$, $x_{3}=-x_{2}\,e^{-i\delta}/\rho$ and $x_{4}=x_{3}^{*}$, where $\rho, \alpha$ and $\delta$ are real parameters.
As compared to the folded or circular string cases, there are now extra degrees of freedom $\rho$ and $\al$, and they correspond to those of $\om_{1}$ and $\om_{2}$ in the profiles.

Another remark is that, in $k\to 1$ limit, the type (i) solutions can be switched to the type (ii) branch without energy costs, changing the number of ``spikes'' ($n$) into the number of crossing the equator ($m$).
This can be understood as changing of the periods for $\mathcal A$- and $\mathcal B$-cycles defined for the two cuts, according to particular modular transformations of the elliptic functions. 
In any case, finite-gap solution interpretation would provide us with a promising framework to test the AdS/CFT from a standpoint of integrable structures of both sides.

\paragraph{Comparison with gauge theory solitons.}

Re-interpreting our solutions as certain coherent states of a spin-chain would be also possible.
It is well-known that folded and circular strings have interpretations as periodic solitons of Landau-Lifshitz equation resulted from string sigma model, which can also be obtained from a coherent-state path-integral of a spin-chain Hamiltonian of gauge theory \cite{Kruczenski03}.
It is also pointed out in \cite{TALK1, TALK2} that a dyonic giant magnon can be compared to a pulse-like soliton in an infinite spin-chain.
There is a localized soliton solution in Landau-Lifshitz model on an infinite line (corresponding to the infinite spin of a dyonic giant magnon), whose dispersion relation can be shown to match that of a dyonic giant magnon, that is, Eq.\,(\ref{Q-mag dispersion relation}) expanded in powers of $\lam/J_{2}^{2}$ (with an identification $Q=J_2$).
It would be then natural to expect our helical solutions with generic $k$ and $v$ also have a direct interpretation via a coherent spin-chain state picture.

\paragraph{Three-spin generalization via soliton technique.}

In this paper, we only cared about the O(4) nonlinear sigma model on {\AdSS} background, and related it to CsG system via Pohlmeyer's reduction.
However, as shown in \cite{Pohlmeyer76, PR79}, one can actually reduce the whole O(6) sigma model to a sort of generalized sine-Gordon like system following similar steps. 
Hence, it would be interesting to investigate the relation between solitons of the generalized sine-Gordon like system and the corresponding string solutions.

\paragraph{}
We hope to revisit these interesting issues as another publication in the near future.

\subsection*{Acknowledgments}

The authors would like to thank H.-Y.~Chen, N.~Dorey, D.~Hofman, Y.~Imamura and B.~Vicedo for their valuable comments and discussions.
KO would like to thank the organizers of the Albert Einstein Institut workshop on Integrability in Gauge and String Theory for a stimulating atmosphere.

\vspace{3mm}

\appendix
\section*{Appendices}

\section{Definitions and Identities for Elliptic Functions\label{app:Elliptic Functions}}

Our conventions for the elliptic functions, elliptic integrals are presented below.

\paragraph{Elliptic theta functions.}
Let $Q=\prod\limits_{n = 1}^\infty  {\left( {1 - e^{2\pi in\tau } } \right)}$. 
We define elliptic theta functions by
\begin{eqnarray}
\vartheta _0 \left( {z,\tau } \right) &\defeq& Q\prod\limits_{n = 1}^\infty  {\left( {1 - 2 \, e^{\pi i (2n - 1) \tau } \cos (2\pi nz) + e^{2\pi i (2n - 1) \tau } } \right)}\,,  \\
\vartheta _1 \left( {z,\tau } \right) &\defeq& 2 \, Q \, e^{i \pi \tau/4} \sin (2\pi z)\prod\limits_{n = 1}^\infty  {\left( {1 - 2 \, e^{2\pi in\tau } \cos (2\pi nz) + e^{4\pi in\tau } } \right)}\,,  \\
\vartheta _2 \left( {z,\tau } \right) &\defeq& 2 \, Q \, e^{i \pi \tau/4} \cos (2 \pi z) \prod\limits_{n = 1}^\infty  {\left( {1 + 2 \, e^{2\pi in\tau } \cos (2\pi nz) + e^{4\pi in\tau } } \right)}\,,  \\
\vartheta _3 \left( {z,\tau } \right) &\defeq& Q \prod\limits_{n = 1}^\infty  {\left( {1 + 2 \, e^{\pi i (2n - 1) \tau } \cos (2\pi nz) + e^{2\pi i (2n - 1) \tau } } \right)}\,.
\end{eqnarray}
We also use an abbreviation $\vartheta _\nu ^0  \equiv \vartheta _\nu  (0,k)$.
The following functions are known as Jacobian theta and zeta functions, respectively:
\begin{equation}
\Theta _\nu  \left( {z,k} \right) \equiv \vartheta _\nu  \left( {\frac{z}{2\eK},\, \tau  = \frac{{i\eK'}}{\eK}} \right),\qquad
Z_\nu  \left( {z,k} \right) \equiv \frac{{\partial _z \Theta _\nu  \left( {z,k} \right)}}{{\Theta _\nu  \left( {z,k} \right)}}\,.
\end{equation}

\paragraph{Complete elliptic integrals.}
Complete elliptic integral of the first kind and its complement are defined as, respectively,
\begin{equation}
\eK(k) \defeq \int_0^1 \frac{dz}{\sqrt{(1-z^2)(1-k^2 z^2)}} \,,\qquad
\eK'(k) \defeq \eK (\sqrt{1 - k^2}) \,.
\end{equation}
We often write $\eK(k)$ as $\eK$. 
Likewise, we omit the moduli parameter $k$ of other elliptic functions or elliptic integrals as well.
There are alternative expressions for $\eK$ and $\eK'$ in terms of elliptic theta functions\,:
\begin{equation}
\eK(k) = \frac{\pi (\vartheta_3 ^0)^2}{2}\,, \qquad
\eK'(k) = - i \eK \tau = \frac{\pi i \tau (\vartheta_3 ^0)^2}{2}\,.
\end{equation}
Complete elliptic integral of the second kind is defined as 
\begin{equation}
\eE(k) \defeq \int_0^1 \sqrt{\frac{1 - k^2 z^2}{1 - z^2}} dz
= \int_0^\eK \dn^2 (u) du\,.
\end{equation}

\paragraph{Jacobian elliptic functions.}
Jacobian sn, dn and cn functions are defined as
\begin{gather}
\sn(z) \defeq \frac{{\vartheta _3^0 }}{{\vartheta _2^0 }} \, \frac{{\vartheta _1 (w)}}{{\vartheta _0 (w)}} \,,\quad
\dn(z) \defeq \frac{{\vartheta _0^0 }}{{\vartheta _3^0 }} \, \frac{{\vartheta _3 (w)}}{{\vartheta _0 (w)}} \,,\quad
\cn(z) \defeq \frac{{\vartheta _0^0 }}{{\vartheta _2^0 }} \, \frac{{\vartheta _2 (w)}}{{\vartheta _0 (w)}} \,, \label{sncndn_def}
\end{gather}
where $z = \pi \pare{\vartheta_3^0}^2 w = 2 \, \eK w$. In terms of Jacobian theta functions, they can be written as
\begin{gather}
\sn(z) = \frac{{\Theta _3 (0) }}{{\Theta _2 (0) }} \, \frac{{\Theta _1 (z)}}{{\Theta _0 (z)}} \,,\quad
\dn(z) = \frac{{\Theta _0 (0) }}{{\Theta _3 (0) }} \, \frac{{\Theta _3 (z)}}{{\Theta _0 (z)}} \,,\quad
\cn(z) = \frac{{\Theta _0 (0) }}{{\Theta _2 (0) }} \, \frac{{\Theta _2 (z)}}{{\Theta _0 (z)}} \,. \label{sncndn_def2}
\end{gather}
The moduli $k$ and $k'\equiv \sqrt{1-k^{2}}$ are related to the elliptic theta functions by 
\begin{equation}
k \equiv \left( {\frac{{\vartheta _2^0 }}{{\vartheta _3^0 }}} \right)^2 ,\qquad k' \equiv \left( {\frac{{\vartheta _0^0 }}{{\vartheta _3^0 }}} \right)^2\,.  \label{kck_def}
\end{equation}
The Jacobian elliptic functions satisfy the following relations\,:
\begin{equation}
\sn^2 (z,k) + \cn ^2 (z,k) = 1, \qquad k^2 \sn^2 (z,k) + \dn ^2 (z,k) = 1\,.
\end{equation}

\section{Some Details of Calculations\label{app:Details of Calculations}}

In this appendix we will collect some key formulae that would be useful in checking the calculation involving the function
\begin{eqnarray}
\hspace{-5mm} \Xi (X,T,w) &=& \frac{\Theta _1 (X - X_0 - w + w_0)}{\Theta _0 (X - X_0) \ssp \Theta_0 (w - w_0)} \, 
\exp \Big( \ssp Z_0 (w - w_0) (X - X_0) + i \ssp u (T - T_0) \Big), \\[1mm]
\hspace{-5mm}  u^2 &=& U + \dn^2 (w - w_0),
\end{eqnarray}
where $X$, $X_0$, $T$ and $T_0$ are all real. For the moment we assume $w$ and $w_0$ to be purely imaginary.
The degrees of freedom of $(T_{0}, X_{0})$ correspond to the initial values for the phases of $\xi_{j}$, and in what follows, we will set them as zero.
We will also set $w_{0}=0$.

As a preliminary, we shall write down several useful formulae concerning elliptic functions.

\medskip \noindent $\bullet$\ \ 
One can express $Z_0 (z,k)$ in terms of Jacobian dn function and complete elliptic integrals as
\begin{equation}
Z_0(z,k) = \int_0^z \dn^2 (u,k) du - z \, \frac{{\eE}}{{\eK}}\,.
\label{zeta0_1}
\end{equation}

\medskip \noindent $\bullet$\ \ 
By using an addition theorem
\begin{equation}
Z_0 (u + v) = Z_0 (u) + Z_0 (v) - k^2 \sn(u)\sn(v)\sn(u + v) \,,
\end{equation}
one can verify the following identities\,:
\begin{align}
\frac{1}{2} \Big( {Z_1 (x + y) + Z_1 (x - y)} \Big) &= Z_0 (x) + \frac{{\sn x \cn x \dn x}}{{\sn^2 x - \sn^2 y}} \,, \\[2mm]
\frac{1}{2} \Big( {Z_1 (x + y) - Z_1 (x - y)} \Big) &= Z_0 (y) - \frac{{\sn y \cn y \dn y}}{{\sn^2 x - \sn^2 y}} \,.
\end{align}

\medskip \noindent $\bullet$\ \ 
Concerning the absolute value of $\Xi (X,T,w)$, one can show that
\begin{equation}
\frac{\Theta _1 (z - w) \, \Theta _1 (z + w)}{\Theta _0 ^2 (z) \, \Theta _0 ^2 (w)} = \frac{k}{{\Theta _0 ^2 (0)}} \, \Big( \sn^2 z - \sn^2 w \Big) \,.
\label{norm_id1}
\end{equation}

\noindent 
With the help of those formulae, we can easily arrived at the following relations\ssp:
\begin{gather}
\left| {\frac{{\partial_X \Xi}}{\Xi}} \right|^2  = \frac{\sn^2(X) \cn^2(X) \dn^2 (X) - \sn^2(w) \cn^2(w) \dn^2 (w)}{\left( \sn^2(X) - \sn^2(w) \right)^2 } \,, \\[2mm]
\Re \!\pare{ \frac{\partial_T \Xi^*}{\overline \Xi} \, \frac{\partial_X \Xi}{\Xi} } = - i u \, \frac{ \sn (w) \cn (w) \dn (w) }{\sn^2(X) - \sn^2 (w)}\,,
\\[2mm]
\Im \!\pare{\frac{\partial_X \Xi}{\Xi}} = \frac{1}{i} \frac{ \sn(w)  \cn(w)  \dn (w) }{\sn^2(X) - \sn^2 (w)}\,,
\end{gather}
which should be useful in evaluating the consistency condition, Virasoro conditions and conserved charges in the main text.

We can now discuss a generalization of the Ansatz \eqref{ansatz xi}. 
In order for $\Xi (X,T,w)$ to be normalizable for all range of $X$, $Z_0 (w, k)$ must be purely imaginary.
When $k$ is real, this can be achieved if and only if $w = m \eK + \iom$ with $m \in \bb{Z}$ and $\omega \in \bb{R}$. 
Therefore, with the Ansatz (\ref{ansatz xi}), general solutions of \eqref{reduced_eom} are given by
\begin{alignat}{2}
\Xi^0 &=   \frac{{\Theta _1 (X  - i\omega )}}{{\Theta _0 (X ) \ssp \Theta_0 (\iom)}} \, 
\exp \Big( {Z_0 (i\omega )X + i u T} \Big) , &\hspace{4mm}
u^2 &= U + \dn^2 (\iom)\,,
\label{Xi-0} \\[2mm]
\Xi^1 &=   \frac{{\Theta _0 (X  - i\omega )}}{{\Theta _0 (X ) \ssp \Theta_1 (\iom)}} \,
\exp \Big( {Z_1 (i\omega )X + i u T} \Big) , &\hspace{4mm}
u^2 &= U - \frac{\cn^2 (\iom)}{\sn^2 (\iom)}\,,
\label{Xi-1} \\[2mm]
\Xi^2 &=   \frac{{\Theta _3 (X  - i\omega )}}{{\Theta _0 (X ) \ssp \Theta_2 (\iom)}} \,
\exp \Big( {Z_2 (i\omega )X + iuT} \Big) , &\hspace{4mm}
u^2 &= U - \frac{(1 - k^2) \sn^2 (\iom)}{\cn^2 (\iom)}\,,
\label{Xi-2}  \\[2mm]
\Xi^3 &=   \frac{{\Theta _2 (X  - i\omega )}}{{\Theta _0 (X ) \ssp \Theta_3 (\iom)}}\exp \Big( Z_3 (i\omega ) X+ i uT \Big) , &\hspace{4mm}
u^2 &= U + \frac{1-k^2}{\dn^2 (\iom)}\,.
\label{Xi-3}
\end{alignat}
These four functions are mutually related by a shift of $w$ as
\begin{alignat}{2}
\Xi^0(X,T;w) &= \Xi(X,T;w = \iom ) \,,&\quad 
\Xi^1(X,T;w) &= - \Xi(X,T;w = \iom - i\eK' ) \,,\cr
\Xi^2(X,T;w) &= \Xi(X,T;w = \iom - \eK-i\eK' ) \,,&\quad 
\Xi^3(X,T;w) &= \Xi(X,T;w = \iom -\eK ) \,. \label{Xi} 
\end{alignat}
Note that in $\omega \to 0$ limit, the functions $\Xi^{0}$, $\Xi^{2}$ and $\Xi^{3}$ reduce to $\sn(X)$, $\dn(X)$ and $\cn(X)$
with the angular velocity satisfying $u^2 = U + 1$, $U$ and $U + 1-k^{2}$, respectively.

It would also be useful to note the properties of $\Xi^i$ given in \eqref{Xi}.
They are doubly periodic with respect to $w$\,:
\begin{equation}
\Xi^i  \to  - \Xi^i \quad \left( {w \to w + 2\eK} \right),\qquad
\Xi^i  \to \Xi^i \quad \left( {w \to w + 2i\eK'} \right),
\end{equation}
and quasi-periodic with respect to $X $\,:
\begin{alignat}{2}
\Xi^0 (X  + 2\eK) &=  - e^{2Z_0 (w)\eK } \, \Xi^0 (X) \,, &\qquad
\Xi^1 (X  + 2\eK) &= e^{2Z_1 (w)\eK } \, \Xi^1 (X) \,, \cr
\Xi^2 (X  + 2\eK) &= e^{2Z_2 (w)\eK } \, \Xi^2 (X) \,, &\qquad
\Xi^3 (X  + 2\eK) &= - e^{2Z_3 (w)\eK } \, \Xi^3 (X) \,.
\label{Xi_per_sig}
\end{alignat}

\end{document}